\documentclass[11pt,a4paper]{article}

\usepackage[no-natbib-sort]{my-jheppub}

\usepackage{graphicx}
\usepackage[english]{babel}
\usepackage{amsthm}

\usepackage[nodayofweek]{date time}

\def \ed  {\end{document} }

\newtheorem{myprop}{Proposition}

\begin{document}

\date{\currenttime}


\title{Off-Shell Spinor-Helicity Amplitudes\\
 from Light-Cone Deformation Procedure}

%
%
%
%
%

\author{Dmitry Ponomarev\footnote{d.ponomarev@imperial.ac.uk}}

\affiliation{Theoretical physics group, Blackett Laboratory, Imperial College London,   SW7 2AZ, U.K.}


\abstract{
We study the consistency conditions for interactions of massless fields of any spin in four-dimensional
flat space using the light-cone approach. We show that they can be equivalently rewritten
as the Ward identities for the off-shell light-cone amplitudes  built from the light-cone
Hamiltonian in the standard way.
Then we find a general solution of these Ward identities.  The solution admits a compact
representation when written in the spinor-helicity form and is given by 
an arbitrary function of spinor products, satisfying  well-known homogeneity constraints.
Thus, we show that the light-cone consistent deformation procedure  inevitably leads to a certain
off-shell version of the spinor-helicity approach.
We  discuss 
 how the relation between the two approaches can be employed  to facilitate the search of consistent interaction of
 massless higher-spin fields.}

\unitlength = 1mm

\today
\date {}
\begin{flushright}\small{Imperial-TP-DP-2016-{03}}\end{flushright}

\maketitle

\def \lR {L}

\def \ll {{\cal \ell}}

\def \be {\begin{equation}}\def \ee {\end{equation}}

\def \ads {AdS$_4$\ }
\def \iffa {\iffalse} 

\def \ed {\bibliography{lcsh}{}
\bibliographystyle{utphys} \end{document}}

\section{Introduction}

Construction of  consistent interacting field theories is an old and challenging problem. 
At the free level elementary fields in the Minkowski space can be identified with unitary irreducible representations
of the Poincare group, which have been classified long time ago \cite{Wigner:1939cj}, for a review see e.g. \cite{Bekaert:2006py}. 
For interacting theories constructed perturbatively the main consistency requirement is that 
the physical degrees of freedom defined at the free level interact without breaking Poincare invariance.

A natural way to construct Poincare invariant theories is to use Lorentz tensors. 
Having contracted tensor indices appropriately one automatically ensures Lorentz
invariance of the action. If, moreover, the action does not depend on coordinates explicitly, then
it is also translationally invariant. In this way, Lorentz tensors allow to make all Poincare
symmetry of the theory manifest.

What makes the manifestly covariant approach much less trivial is that, unless special care is taken,
it introduces unwanted degrees of freedom. For consistency these extra degrees of freedom should be removed
from the theory, usually, either  by constraints or as a result of gauge invariance.
In particular, for massless fields with spin greater than one-half description in terms
of Lorentz tensors requires these fields to be gauge ones.
 To avoid unwanted degrees of freedom at the 
interacting level gauge invariance should  also be preserved. This
leads to the manifestly Lorentz covariant deformation procedure, which 
amounts to a simultaneous deformation of the action and gauge transformations in a way that
the action remains gauge invariant.
For
massless higher-spin fields this approach was used by many authors and
leads to the conclusion that non-trivial local interactions  of higher-spin fields in flat space cannot exist, see e.g. \cite{Aragone:1979hx,Bekaert:2010hp,Joung:2013nma}\footnote{There are also other types of no-go arguments, e.g.  \cite{Weinberg:1964ew,Coleman:1967ad}.
For a comprehensive review
on no-go theorems for massless higher-spin interactions and how they can be circumvented,  see \cite{Bekaert:2010hw}.}.

An alternative approach is to abandon Lorentz tensors and control Poincare symmetry explicitly
by requiring that the Noether charges, deformed
order by order, obey commutation relations of the Poincare algebra.
A particular version of this procedure is the light-cone approach \cite{Bengtsson:1983pd,Bengtsson:1983pg,Bengtsson:1986kh},
 which we will use in this paper.
 A somewhat unattractive feature of the light-cone approach is that
it requires manual control of all symmetries and as a result at first sight  appears  less economical than the manifestly
 Lorentz covariant one. On the other hand, it has an advantage of being completely general. It is quite remarkable
that already at the cubic order
 by giving up manifest Lorentz covariance one can find additional consistent local interactions\footnote{At least formally, the exotic
 vertices can be written in the Lorentz covariant form, but this requires non-localities \cite{Conde:2016izb,Sleight:2016xqq}.}. These
 exotic vertices are known for a long time \cite{Bengtsson:1986kh,Metsaev:1991mt,Metsaev:1991nb,Metsaev:1993ap},
  but  only recently it was emphasised that they
 are missing in the manifestly Lorentz covariant classification 
 \cite{Bengtsson:2014qza}\footnote{Note that within the manifestly covariant framework there exist some additional lower-derivative deformations of the gauge algebra and gauge transformations, that cannot be promoted to the action  level, see e.g. \cite{Boulanger:2006gr,Bekaert:2010hp}. It
 would be interesting to see whether they are related to the exotic vertices from the light-cone approach.}.
  Moreover, presence of the exotic vertices turns out to be crucial for consistency of  
   higher-spin interactions at the quartic order \cite{Metsaev:1991mt,Metsaev:1991nb}.
 In particular, they are present in the chiral higher-spin theory \cite{Ponomarev:2016lrm}, which is a cubic theory consistent to all orders in interactions,
 see also \cite{Metsaev:1991mt,Metsaev:1991nb}.

In this paper we study the light-cone consistency conditions in four-dimensional flat space-time  and show that they can be
rephrased as the Ward identities for the off-shell  light-cone 
amplitudes.
This terminology deserves clarification.
Firstly, a term 'Ward identity' is usually used for constraints on the $S$-matrix, imposed by gauge invariance.
In the light-cone approach gauge freedom is completely fixed and  the constraints that we call  'Ward identities'
appear as a consequence of invariance of the $S$-matrix with respect to  the global Poincare
symmetry algebra. Nevertheless, we then show that these constraints are equivalent  to the constraints imposed by gauge invariance
in manifestly covariant approaches, thus justifying a name 'Ward identities' that we use for them.
 Secondly, by an 'off-shell amplitude' we, essentially, mean an amputated correlator, that is a correlator where
 propagators associated with external lines are removed, but  external momenta are not put on-shell.
In the following, these light-cone amplitudes will be built
from the light-cone Hamiltonian according to the light-cone Feynman rules to be specified in the text.

Next, we solve these Ward identities and show that their 
general solution  can be compactly written employing the spinor-helicity language as
an arbitrary function of spinor contractions, obeying natural homogeneity constraints\footnote{Following the higher-spin literature we
  use 'spinor-helicity'
just to term a representation for the total amplitude when it is given as a function of spinor products.
 In contrast, originally it rather means a set of tools employed
to evaluate QCD Feynman diagrams, which include: the spinor-helicity representation for the polarisation vectors, prescriptions
for a convenient choice of reference spinors, factorisation of external momenta in terms of spinors, etc.}. 
  The relation
between the light-cone deformation procedure and the spinor-helicity approach was observed before. In \cite{Ananth:2012un,Akshay:2014qea}
it was found that the cubic light-cone interactions can be easily rewritten in the spinor-helicity form. 
The associated three-point amplitudes reproduce those found in \cite{Benincasa:2007xk}.
Then, in
\cite{Bengtsson:2016jfk,Bengtsson:2016alt} the connection with the spinor-helicity approach was used to study the quartic sector of the light-cone
consistency equations\footnote{In a slightly different context the relation between the light-cone gauge and the spinor-helicity approach was found in
\cite{Chalmers:1998jb}. Namely, 
 the spinor-helicity approach was identified as fixing the space-cone gauge,
which is closely related to  the light-cone one.}.

In the present paper we show that the relation between the light-cone deformation procedure and the spinor-helicity representation is not
 accidental. Instead, spinor contractions and associated homogeneity constraints appear inevitably from the
light-cone consistency conditions. Moreover, being essentially off-shell 
 the light-cone approach provides a natural off-shell continuation of  amplitudes
 written in the spinor-helicity representation.
 In this way,  by combining benefits of the two approaches
we obtain
 an attractive tool to study consistent theories of massless fields:  on one hand, it is manifestly
 covariant, which makes it efficient and, on the other hand, it is completely general and hence  captures all 
 consistent local interactions.

The plan of the paper is as follows. In Section \ref{sec7oct1} we review the basics of the light-cone deformation
procedure. We start with the free theory and show how requiring closure of the Poincare algebra one derives
kinematical and dynamical constraints. In Section \ref{section5oct1} we show that the dynamical constraints
can be equivalently rewritten as the Ward identities for the light-cone amplitude.
To this end, in Section \ref{sec29oct1} we first rewrite them in terms of the light-cone Hamiltonian only. 
In Section \ref{sect2oct1} we collect some extra notations and identities necessary to streamline the following analysis.
Next,
in Section \ref{29oct3} we show
that after some manipulations the light-cone consistency conditions acquire a form of the Ward identities for the amplitude, that 
contains contact diagrams, exchanges with one internal line and extra terms bilinear in the 
light-cone Hamiltonian. After that, in Section \ref{sec29oct2} we show that these extra terms 
produce contributions of exchanges with two internal lines as well as other terms, which are
cubic in the light-cone Hamiltonian. Applying this procedure iteratively we recover the 
Ward identity for the complete amplitude in Section \ref{29oct4}.
 This result is
summarised in Proposition  \ref{prop2oct1}.  Then, in 
Section \ref{sect1oct1} we derive a general solution of the Ward identity. The solution is summarised in 
Proposition \ref{prop5oct1}.  Interpretation of these results in terms of the spinor-helicity approach
 is given
in Section \ref{sect5oct3}. We present our conclusions and discuss possible extension in Section \ref{sec7oct2}.
Appendix \ref{ap3oct1} summarises our conventions.

\section{Basics}
\label{sec7oct1}

In this Section we review the basics of the light-cone deformation procedure
for massless fields in four-dimensional flat space-time.

\subsection{Free massless fields in the light-cone gauge}

The Poincare algebra commutation relations are given by
\begin{align}
\notag
[P^a,P^b] &\,= 0,\\
\notag
[J^{ab},P^c] &\,= P^a \eta^{bc}- P^b \eta^{ac},\\
[J^{ab}, J^{cd}] &\, = J^{ad} \eta^{bc} -J^{bd} \eta^{ac} -J^{ac} \eta^{bd}+ J^{bc} \eta^{ad},
\end{align}
where $x^a=\{x^-,x^+,x,\bar x\}$. For more details on our conventions see Appendix \ref{ap3oct1}.
 It admits helicity-$\lambda$ representations 
\begin{align}
\notag
P^a \cdot \Phi^{\lambda}&\,\equiv \partial^a \Phi^{\lambda},\\
\label{29sep2}
 J^{ab}\cdot \Phi^{\lambda}&\,\equiv (x^a\partial^b - x^b \partial^a+S^{ab} )\Phi^\lambda,
\end{align}
where $S^{ab}$ is the spin part of the angular momentum. In the light-cone gauge
\begin{equation}
\label{29sep4}
S^{+a}\cdot \Phi^{\lambda} =0,\qquad  S^{ab} \partial_a \cdot \Phi^{\lambda}=0.
\end{equation}
The first condition in (\ref{29sep4}) implies that the only non-vanishing components of $S^{ab}$ are $S^{x\bar x}$, $S^{x-}$
and $S^{\bar x -}$. The second condition allows to express all of them in terms of $S^{x\bar x}$
\begin{equation}
\label{21sep3}
S^{x-}\cdot \Phi^{\lambda}= -S^{x\bar x}\cdot \frac{\partial}{\partial^+}\Phi^{\lambda}, \qquad 
S^{\bar x -}\cdot \Phi^{\lambda}= S^{x\bar x}\cdot  \frac{\bar\partial}{\partial^+}\Phi^{\lambda}.
\end{equation}
Thus, the helicity representation is specified by the action of $S^{x\bar x}$
generating  the Wigner little group. It is conventional to define
\begin{equation}
\label{29sep3}
S^{x\bar x} \cdot \Phi^{\lambda} = -\lambda \Phi^{\lambda},
\end{equation}
where $\lambda$ is the helicity.

\paragraph{Free action and canonical analysis.}
The canonically normalised action for the set of free massless fields  is
given by
\begin{equation}
\label{19sep6}
S_2\equiv \int d^4 x L_2, \qquad L_2 = -\frac{1}{2}\sum_{\lambda}  \partial_a\Phi^{-\lambda} \partial^a \Phi^\lambda.
\end{equation}
Here we do not impose any restrictions on the spectrum of values of $\lambda$ except that opposite helicities
should enter together. For example, for the spin $s$ field one has $\lambda = \{s,-s\}$ and
\begin{equation}
L_2= -\partial_a\Phi^{-s}\partial^a \Phi^s.
\end{equation}
In the scalar case $\lambda=0$ and 
\begin{equation}
L_2 = -\frac{1}{2} \partial_a\Phi^{0} \partial^a \Phi^0.
\end{equation}

In the light-cone coordinates $\partial^-$ is the time derivative,  so the canonical momentum is 
 \begin{equation}
 \Pi^\lambda \equiv \frac{\delta L_2}{\delta( \partial^- \Phi^\lambda)}= -\partial^+ \Phi^{-\lambda}.
 \end{equation}
Then, the Poisson bracket is
  \begin{equation}
 [\partial^+ \Phi^\lambda(x^\perp,x^+),\Phi^{\mu}(y^\perp,x^+)]_P= \delta^{\lambda+\mu,0}\delta^{3}(x^\perp,y^\perp),
 \end{equation}
 where we use $x^{\perp}\equiv \{x,\bar x,x^- \}$.
 The canonical Hamiltonian is given by
 \begin{equation}
 \label{21sep8}
  H_2\equiv  \sum_{\lambda }  \int  d^{3}x^\perp(\Pi^\lambda \partial^-\Phi^{\lambda} - L_2)  =
  \sum_{\lambda} \int d^{3}x^\perp \partial \Phi^{-\lambda} \bar\partial \Phi^\lambda,
 \end{equation}
 where one integrates over equal-time hypersurfaces.

It is not hard to see that due to the fact that the Lagrangian (\ref{19sep6}) is first order in time derivatives, the
theory features constraints. Their analysis has been discussed by many authors, for a review see e.g. \cite{Heinzl:2000ht}.
As a result, to account for the constraints one should replace the Poisson bracket 
by the Dirac one
 \begin{equation}
  \label{21sep6}
 [\partial^+ \Phi^\lambda(x^\perp,x^+),\Phi^{\mu}(y^\perp,x^+)]= \frac{1}{2}\delta^{\lambda+\mu,0}\delta^{3}(x^\perp,y^\perp),
 \end{equation}
or, equivalently,
  \begin{equation}
  \label{21sep7}
 [\Phi^\lambda(x^\perp,x^+),\Phi^{\mu}(y^\perp,x^+)]= \frac{1}{\partial^+_x - \partial^+_y}\delta^{\lambda+\mu,0}\delta^{3}(x^\perp,y^\perp).
 \end{equation}
 
 The canonical Hamiltonian (\ref{21sep8}) and the Dirac bracket (\ref{21sep7}) define the time evolution 
 \begin{equation}
 \label{19sep11}
 \partial^- F(\Phi) = [F(\Phi), H_2].
 \end{equation}
In particular, one can verify that 
 \begin{equation}
\partial^-\Phi^\lambda(x) = [\Phi^\lambda(x), H_2] = -\frac{\partial \bar \partial}{\partial^+} \Phi^\lambda(x),
 \end{equation}
which is consistent with the variation of (\ref{19sep6}).

 \paragraph{Noether currents and charges.} 
 The action (\ref{19sep6}) is invariant with respect to transformations (\ref{29sep2}).
 The associated Noether currents are well known
  \begin{align}
\notag
 P^i \quad \to  \;\quad T^{i,j} &\,= \sum_{\lambda}\frac{\delta L_2}{\delta (\partial_j \Phi^\lambda)}\partial^i \Phi^\lambda - \eta^{ij} L_2,\\
 J^{ij} \quad   \to  \quad L^{ij,k} &\,= x^i T^{j,k}-x^j T^{i,k}+R^{ij,k},
  \label{21sep10}
 \end{align}
  where $R^{ij,k}$ is the spin current
  \begin{equation}
  \label{21sep12}
  R^{ij,k} = \sum_{\lambda}\frac{\delta L_2}{\delta( \partial_k \Phi^\lambda)} S^{ij}\cdot \Phi^\lambda
  \end{equation}
 and  $S^{ij}$ was given in (\ref{21sep3}), (\ref{29sep3}).
  
  Accordingly, we define the Noether charges 
    \begin{equation}
  \label{21sep13}
  P^i_2 = \int d^3x^\perp T^{i,+}, \qquad J^{ij}_2 = \int d^3x^\perp L^{ij,+},
  \end{equation}
  which with a slight abuse of notations we denoted by the same
  symbols as the algebra generators themselves.
  For simplicity we  choose the integration hypersurface to be $x^+=0$.
  Explicitly the charges (\ref{21sep13}) read 
  \begin{equation}
  \label{21sep14}
  P^i_2 = -\sum_{\lambda}\int d^3x^\perp \partial^+\Phi^{-\lambda}  p_2^i \Phi^\lambda, \qquad  J_2^{ij} = -\sum_{\lambda}\int d^3x^\perp \partial^+\Phi^{-\lambda}  j_2^{ij} \Phi^\lambda,
  \end{equation}
  where 
  \begin{align}
\notag
p_2^+ &\,=\partial^+, & p_2^- &= -\frac{\partial \bar\partial}{\partial^+}, \qquad\qquad \quad\;  p_2 = \partial,  \qquad\qquad\quad\;    \bar p_2=
\bar\partial,\\
\notag
j_2^{+-} &\,=  - x^-\partial^+,&  j_2^{x\bar x} &= x\bar\partial - \bar x \partial -\lambda, \\
\notag
j_2^{x+}&\, = x\partial^+, &  j_2^{x-} &= -x\frac{\partial \bar\partial}{\partial^+} - x^- \partial+\lambda\frac{\partial}{\partial^+}, \\
j_2^{\bar x+}&\, = \bar x\partial^+,
 &  j_2^{\bar x-} &= -\bar x\frac{\partial \bar\partial}{\partial^+} - x^- \bar\partial
-\lambda\frac{\bar\partial}{\partial^+}.
\label{21sep15}
\end{align}
This representation coincides with the original one (\ref{29sep2}) up to terms that vanish on $x^+=0$ and on the mass-shell.
As expected,  the charges (\ref{21sep14}) generate the algebra action via the commutator
  \begin{equation}
  \label{21sep16}
 [\Phi^\lambda, P_2^i] = p_2^i \Phi^\lambda, \qquad  [\Phi^\lambda, J_2^{ij}] = j_2^{ij} \Phi^\lambda.
  \end{equation} 
 Moreover, the charge $P^-_2$ associated with the light-cone time translation is the canonical Hamiltonian
 $H_2$  (\ref{21sep8}).

 \paragraph{Fourier transform.} It is convenient to make the Fourier transform with respect to spatial coordinates 
 $x^-$, $x$ and $\bar x$
  \begin{align}
\notag
  \Phi(x,x^+) &\,= (2\pi)^{-\frac{3}{2}} \int{e^{+i (x^- p^+ + \bar x p +x\bar p)}\Phi(p,x^+)d^{3}p^\perp},\\
    \label{21sep17}
   \Phi(p,x^+) &\,= (2\pi)^{-\frac{3}{2}} \int{e^{-i (x^- p^+ + \bar x p +x\bar p)}\Phi(x,x^+)d^{3}x^\perp},
  \end{align}
followed by a change of variables $p=iq$. This allows to avoid complex factors and effectively amounts to the  substitution
  \begin{equation}
  \label{21sep19}
  \frac{\partial}{\partial x^i} \to q_i, \qquad x^i \to -\frac{\partial}{\partial q_i}.
  \end{equation}
We will also  use $q^\perp \equiv \{q,\bar q,q^+\}$ and $\beta \equiv q^+ $.

In these terms the canonical commutator reads
\begin{equation}
\label{29sep5}
[\Phi^{\lambda_1}(q^\perp_1,x^+),\Phi^{\lambda_2}(q^\perp_2,x^+)] = \frac{\delta^{\lambda_1+\lambda_2,0}\delta^3(q^\perp_1+q^\perp_2)}{\beta_1-\beta_2}
\end{equation}
and the Noether charges are 
\begin{align}
\notag
P_2^i &\,= \sum_{\lambda}\int d^3q^\perp_1 d^3q^\perp_2 \delta^3(q^\perp_1+q^\perp_2)\beta_1 \Phi^{-\lambda}(q^\perp_1,x^+) p^i_2(q_2,\partial_2) \Phi^\lambda(q^\perp_2,x^+),\\
\label{29sep6}
J_2^{ij} &\,= \sum_{\lambda}\int d^3q^\perp_1 d^3q^\perp_2 \delta^3(q^\perp_1+q^\perp_2)\beta_1 \Phi^{-\lambda}(q^\perp_1,x^+) j^{ij}_2(q_2,\partial_2) \Phi^\lambda(q^\perp_2,x^+),
\end{align}
where
  \begin{align}
\notag
p_2^+ &\,=q^+, & p_2^- & = -\frac{q \bar q}{\beta}, \qquad\qquad \quad\; p_2 = q, \qquad\qquad \quad\; \bar p_2=
\bar q,\\
\notag
j_2^{+-} &\,=  \frac{\partial}{\partial \beta}\beta, &  j_2^{x\bar x} &=  N_q- N_{\bar q}  -\lambda,\\
\notag
j_2^{x+}&\, = -\beta \frac{\partial}{\partial \bar q}, &  j_2^{x-} &= \frac{\partial}{\partial \bar q}\frac{q \bar q}{\beta} + q \frac{\partial}{\partial \beta}+\lambda\frac{q}{\beta},\\
j_2^{\bar x+}&\, = -\beta \frac{\partial}{\partial q},
 & j_2^{\bar x-} &= \frac{\partial}{\partial q}\frac{q \bar q}{\beta} +  \bar q \frac{\partial}{\partial \beta}
-\lambda\frac{\bar q}{\beta}
\label{29sep7}
\end{align}
and 
\begin{equation}
N_q \equiv q\frac{\partial}{\partial q}, \qquad N_{\bar q} \equiv \bar q\frac{\partial}{\partial \bar q}.
\end{equation}

\subsection{Introducing interactions consistently}

At the interacting level the action receives non-linear corrections and so do the charges (\ref{29sep6}).
The only consistency requirement that one imposes is that they still generate the Poincare algebra.
The standard lore of the light-cone approach is that it is sufficient to deform only the generators,
that are transversal to the light-cone $x^+=0$ \cite{Dirac:1949cp}. These are
\begin{equation}
H\equiv P^- ,\qquad J\equiv J^{x-}, \qquad \bar J\equiv J^{\bar x-} 
\end{equation}
and they are called dynamical generators\footnote{In fact, one can even succeed by deforming
only $P^-$, but then this deformation rather plays a role of the amplitude than of the Hamiltonian, as it will be discussed later, i.e.
see Proposition \ref{prop2oct1}.}. The remaining generators are called kinematical.
Let us collectively denote  by $D$ and $K$ the charges of the dynamical and the kinematical generators 
respectively. Then, at the non-linear level
\begin{equation}
D =D_2 + \delta D, \qquad K = K_2.
\end{equation}

Accordingly, one can break the Poincare algebra commutation relations into classes depending
on the types of generators they feature. The simplest type of commutators is 
\begin{equation}
[K,K]=K,
\end{equation}
as it is automatically satisfied at the non-linear level.
 Other two groups of commutators 
 \begin{align}
[K,D] = K \qquad &\Rightarrow \qquad [K,\delta D] = 0,\\
[K,D] = D \qquad &\Rightarrow \qquad [K,\delta D] = \delta D
\end{align}
are very similar to each other.
They both result in linear differential equations on deformations $\delta D$, which will be called kinematical
constraints. These constraints can be solved easily for $\delta D$ at any order of deformation.
The last type of commutators is
\begin{equation}
[D,D]=0
\end{equation}
and it presents the main difficulty of the light-cone deformation procedure. We will now consider these
issues one by one in more details.

\paragraph{Deformation.}
A general ansatz for the dynamical generators at the non-linear level is 
\begin{equation}
H=H_2+\sum_n H_n ,\quad J=J_2+\sum_n J_n, \quad \bar J=\bar J_2+\sum_n \bar J_n,
\end{equation}
where
 \begin{align}
 \notag
  H_n  =&\;  \frac{1}{n!}\sum_{\lambda_i}\int d^{3n}q^\perp \delta^3 (\sum_{i=1}^n q^\perp_i) h^{\lambda_1 \dots \lambda_n}_n \prod_{i=1}^n\Phi^{\lambda_i}(q^\perp_i),\\
\notag
  J_n  =&\;   \frac{1}{n!}\sum_{\lambda_i}\int d^{3n}q^\perp \delta^3 (\sum_{i=1}^n q^\perp_i) \Big[j_n^{\lambda_1 \dots \lambda_n} -\frac{1}{n}
  h_n^{\lambda_1\dots \lambda_n}\big( \sum_j \frac{\partial}{\partial \bar q_j}\big)\Big]\prod_{i=1}^n\Phi^{\lambda_i}(q^\perp_i),\\
    \label{22sep1}
  \bar J_n  =&\;  \frac{1}{n!}\sum_{\lambda_i}\int d^{3n}q^\perp \delta^3 (\sum_{i=1}^n q^\perp_i) \Big[\bar j_n^{\lambda_1 \dots \lambda_n} -\frac{1}{n}
  h_n^{\lambda_1\dots \lambda_n}\big( \sum_j \frac{\partial}{\partial  q_j}\big)\Big]\prod_{i=1}^n\Phi^{\lambda_i}(q^\perp_i).
\end{align}
Here $h_n$, $j_n$ and $\bar j_n$ generalise operators appearing in (\ref{29sep7}).
They all depend on transversal momenta $q^\perp_i$, but  not  on $q_i^-$. On the one hand, $q_i^-$-independence of 
interaction vertices can always be 
achieved by field redefinitions, using the fact that on-shell 
\begin{equation}
q_i^- \approx h_2(q^\perp_i)\equiv -\frac{q_i\bar q_i}{\beta_i}.
\end{equation}
On the other hand, interactions 
 that are free of time derivatives are convenient, as they do not deform the canonical bracket.

In (\ref{22sep1}) the momentum delta-functions 
ensure translation invariance along spatial directions, which implies that the kinematical
constraints arising from commutators with $P_2$, $\bar P_2$ and $P^+_2$ are automatically 
satisfied. The $h$-dependent corrections in
the ansatzes for $J_n$ and $\bar J_n$ in (\ref{22sep1}) is just 
a standard trick which slightly simplifies
 the remaining kinematical constraints for $j$ and $\bar j$.

\paragraph{Kinematical constraints.}
To evaluate the remaining kinematical constraints it
 is convenient to use
\begin{align}
\label{30sep1}
[F(\Phi),J_2^{ij}] &\,= [\Phi,J_2^{ij}] \,\frac{\delta F(\Phi)}{\delta \Phi} = j_2^{ij} \Phi \,\frac{\delta F(\Phi)}{\delta \Phi},
\end{align}
which is a result of consecutive  application of  (\ref{21sep16}) to each $\Phi$ entering $F(\Phi)$.
Employing (\ref{30sep1})
to evaluate commutators with $H_n$, we find
\begin{align}
\label{30sep2}
[J_2^{x+},H_n]=0\qquad \Rightarrow &\qquad \delta^3 (\sum_{i=1}^n q^\perp_i)\sum_{i=1}^n \beta_i \frac{\partial}{\partial \bar q_i} h_n^{\lambda_1\dots \lambda_n}=0,\\
\label{30sep3}
[J_2^{\bar x+},H_n]=0\qquad \Rightarrow &\qquad \delta^3 (\sum_{i=1}^n q^\perp_i)\sum_{i=1}^n \beta_i \frac{\partial}{\partial  q_i} h_n^{\lambda_1\dots \lambda_n}=0,\\
\label{30sep4}
[J_2^{x\bar x},H_n]=0\qquad \Rightarrow &\qquad \delta^3 (\sum_{i=1}^n q^\perp_i)\sum_{i=1}^n (N_{q_i}- N_{\bar{q}_i}+\lambda_i)h_n^{\lambda_1\dots \lambda_n}=0,\\
\label{30sep5}
[J_2^{+-},H_n]+H_n=0\qquad \Rightarrow &\qquad \delta^3 (\sum_{i=1}^n q^\perp_i) \sum_{i=1}^n \beta_i \frac{\partial}{\partial \beta_i} h_n^{\lambda_1\dots \lambda_n}=0.
\end{align}
Constraints for $j_n$ and $\bar j_n$ are analogous and can be found, e.g. in \cite{Ponomarev:2016lrm}.

The first two conditions (\ref{30sep2}), (\ref{30sep3}) imply that $h_n$ can depend on $q_i$ and $\bar q_i$ only through
their particular combinations with $\beta_i$
\begin{equation}
\bar{\mathbb{P}}_{ij} \equiv \bar q_i\beta_j -\bar q_j \beta_i , \qquad \mathbb{P}_{ij} \equiv q_i\beta_j -q_j \beta_i.
\end{equation}
The remaining two conditions (\ref{30sep4}), (\ref{30sep5}) simply specify the homogeneity degrees of 
$h_n$ on its arguments.

  \paragraph{Dynamical commutators.}
  Let us now consider the dynamical equation
  \begin{equation}
  \label{22sep5}
 [H,J]=0 \qquad \Rightarrow \qquad [H_2, J_n] + [H_3, J_{n-1}] + \dots + [H_{n-1}, J_3]+ [H_n,J_2]=0.
  \end{equation}
   The charges $H_2$ and $J_2$ are already known,
   so the first and the last commutators can be
  readily computed. Employing (\ref{30sep1}) for $H_2$ we find
  \begin{align}
  [H_2, J_n]  
  =  -\frac{1}{n!}\sum_{\lambda_i}\int d^{3n}q^\perp \delta (\sum_{i=1}^n q^\perp_i)  
  {\cal H}^{\{i\}}
 \Big[j_n^{\lambda_1 \dots \lambda_n} +\frac{1}{n}
 \big( \sum_j \frac{\partial}{\partial \bar q_j}\big) h_n^{\lambda_1\dots \lambda_n}\Big]
 \prod_{i=1}^n\Phi^{\lambda_i}(q^\perp_i),
  \label{22sep6}
  \end{align}
  where 
  \begin{equation}
  {\cal H}^{\{i\}}\equiv \sum_{i=1}^n h_2(q^\perp_i).
  \end{equation}
 
 Analogously, the last commutator in (\ref{22sep5}) gives
 \begin{align}
  \label{22sep7}
 [H_n,J_2] = \frac{1}{n!}\sum_{\lambda_i}\int d^{3n}q^\perp \delta (\sum_{i=1}^n q^\perp_i){\cal J}^{\{i\}} h_n^{\lambda_1 \dots \lambda_n} 
 \prod_{i=1}^n\Phi^{\lambda_i}(q^\perp_i),
 \end{align}
 where
 \begin{equation}
 \label{22sep8}
{\cal J}^{\{i\}} =\sum_{i=1}^n \Big(-\frac{q_i\bar q_i}{\beta_i}\frac{\partial}{\partial \bar q_i}-q_i \frac{\partial}{\partial \beta_i}+\lambda_i \frac{q_i}{\beta_i}\Big).
 \end{equation}
 
 Eventually, (\ref{22sep5}) becomes
  \begin{align}
\notag
 &\,-\frac{1}{n!}\sum_{\lambda_i}\int d^{3n}q^\perp \delta (\sum_{i=1}^n q^\perp_i)  
  {\cal H}^{\{i\}}
 \Big[j_n^{\lambda_1 \dots \lambda_n} +\frac{1}{n}
 \big( \sum_j \frac{\partial}{\partial \bar q_j}\big) h_n^{\lambda_1\dots \lambda_n}\Big]
 \prod_{i=1}^n\Phi^{\lambda_i}(q^\perp_i)\\
\notag
 &\, \qquad \qquad\qquad\qquad\qquad \;+
 \frac{1}{n!}\sum_{\lambda_i}\int d^{3n}q^\perp \delta (\sum_{i=1}^n q^\perp_i){\cal J}^{\{i\}} h_n^{\lambda_1 \dots \lambda_n} 
 \prod_{i=1}^n\Phi^{\lambda_i}(q^\perp_i)\\
    \label{22sep9}
 &\,\qquad\qquad\qquad\qquad\qquad\qquad\qquad\qquad \qquad \quad+ [H_3, J_{n-1}] + \dots + [H_{n-1}, J_3]=0.
  \end{align}
 
 The equation $[H,\bar J]=0$ is analogous. Moreover, $[H,J] = 0$ and $[H,\bar J]=0$ together imply that
the last consistency condition $[J,\bar J]=0$ is also satisfied \cite{Ponomarev:2016lrm}. Hence, in order  to proceed  it is enough to learn how to solve
(\ref{22sep9})  efficiently.

\section{Towards the Ward identity}
\label{section5oct1}

It is hard not to notice that the light-cone consistency condition (\ref{22sep9}) is reminiscent of some
constraint imposed on the total amplitude made of $H$. Indeed, along with a contribution from the 
contact n-point interaction $h_n$, it contains terms $[H_m,J_{n+2-m}]$, which are naturally associated
with the exchanges involving $m$- and $(n+2-m)$-point vertices. On the other hand, it is not at all obvious
how to make this relation precise. Firstly, (\ref{22sep9}) contains $J$ and $H$ on equal footing. While $H$
is trivially related to the vertices the way they appear in the action, for $J$ this relation is less obvious. Secondly,
it is not immediately clear how (\ref{22sep9}) produces contributions associated with exchanges involving two or more
internal lines.

In the amplitude language, the main consistency requirement any interacting field theory should satisfy 
is that the $S$-matrix is Poincare invariant.
Given that the $S$-matrix is essentially the transition amplitude between the on-shell states of the free theory,
 the Poincare algebra acts on the $S$-matrix by the free theory generators. Of course, one expects that
 consistency conditions in different approaches are related to each other. 
 Hence,  the light-cone consistency condition (\ref{22sep9}) should be related to Poincare invariance of the
$S$-matrix with respect to the  free theory transformations. Our goal in this Section is to clarify this relation. 
This will enable us to rewrite the light-cone consistency conditions in the $S$-matrix-like form and
 then solve them
in Section \ref{sect1oct1}.

More precisely, we will show that 
\begin{align}
\notag
[H,J] = 0 \qquad \Leftrightarrow\qquad  [A,J_2]=0,\\
\label{15oct1}
[H,\bar J] = 0 \qquad \Leftrightarrow \qquad  [A,\bar J_2]=0,
\end{align}
where $A$ will be specified later. At this point we just note that $A$, similarly to $H$, is given by a space-time integral,
 where as a kernel instead of $h$ one has a certain off-shell continuation of the amplitude built of $h$. Moreover, one can 
 then trivially show that
  \begin{equation}
 \label{22oct1}
 [H,K]=0 \qquad \Leftrightarrow \qquad [{A},K]=0,
 \end{equation}
 were $K$ are kinetic generators as well as
 \begin{equation}
 \label{22oct2}
 [{A},H] = {\cal H}\, {A}\approx 0.
 \end{equation}
Combining (\ref{15oct1})-(\ref{22oct2}) together, we find that the light-cone consistency conditions imply Poincare invariance
of the $S$-matrix, as expected. However, we would like to emphasise, that (\ref{15oct1})-(\ref{22oct2}) hold off-shell.
Also, let us stress that despite these formulas will be derived for massless particles in flat four-dimensional space, 
they are, clearly, completely general and should be valid for any number of dimensions, types of particles and the value
of the cosmological constant.

\subsection{Eliminating $J$}
\label{sec29oct1}

What complicates the analysis of (\ref{22sep9}) is that it has to be solved for two unknowns $h_n$ and $j_n$. At the same time, 
conceptually, it is clear that once $h_n$ is known one can find the action, which, if Poincare-invariant, 
  defines $j_n$. So, our first goal is to eliminate $j_n$ in favour of $h_n$.

An important observation, which will be used extensively throughout the paper is that the operator,
that acts on $h_n$ in (\ref{22sep9}) has the following property
\begin{equation}
\label{1oct3}
\big({\cal J}^{\{i\}}-  {\cal H}^{\{i\}}\frac{1}{n}
  \sum_{i=1}^n \frac{\partial}{\partial \bar q_i}\big) (\sum_{i=1}^n q^{\perp}_i)\propto (\sum_{i=1}^n q^{\perp}_i).
\end{equation}
This allows to use momentum conservation inside $h_n$ without changing its contribution to 
(\ref{22sep9}).
On the other hand, since
\begin{equation}
\big[\frac{1}{n}
  \sum_{i=1}^n \frac{\partial}{\partial \bar q_i}, \sum_{j=1}^n \bar q_j\big]=1,
\end{equation}
one can always add to $h_n$ terms proportional to the total momentum
\begin{equation}
\label{23oct1}
h_n^{\lambda_1\dots \lambda_n} \quad \to \quad  \tilde h_n^{\lambda_1\dots \lambda_n}=
h_n^{\lambda_1\dots \lambda_n} + \alpha \sum_{j=1}^n \bar q_j,
\end{equation}
so that 
 \begin{equation}
 \label{22sep10}
 j_n^{\lambda_1 \dots \lambda_n} +\frac{1}{n}
 \big( \sum_j \frac{\partial}{\partial \bar q_j}\big) \tilde h_n^{\lambda_1\dots \lambda_n}=0
 \end{equation}
 is satisfied. In other words, once a solution $h^n$ of (\ref{22sep9}) is found, one can replace it
 with $\tilde h_n$, which additionally satisfies (\ref{22sep10}). Moreover, using the fact that  $h^n$
 enters  $H^n$ multiplied by the momentum conserving delta function, the replacement of
 $h^n$ with $\tilde h^n$ leaves commutators $[H_n,J_m]$ intact as well. In the following we will omit the
 tilde and write just $h_n$. Clearly, a similar argument works for $\bar j_n$.

 The condition (\ref{22sep10}) allows to solve for $j_n$ in terms of $h_n$.
 This leads to
 \begin{align}
 \label{22sep12}
 J_n 
 =-\frac{1}{n!}\sum_{\lambda_i}\int d^{3n}q^\perp d\bar \varepsilon \dot\delta(\bar\varepsilon)  \delta(\sum_{i=1}^n q^\perp_i+ \bar\varepsilon)
 h_n^{\lambda_1 \dots \lambda_n}  \prod_{i=1}^n\Phi^{\lambda_i}(q^\perp_i),
 \end{align}
 where we found convenient to introduce an extra variable $\bar\varepsilon$ to write a derivative of the momentum conserving delta-function
 in a more concise and symmetric form. Comparing it with (\ref{22sep1}) we find that this derivative is the only difference between $H_n$
 and $J_n$.

 This has the following simple interpretation. Given that at the non-linear level from
 all $P^i$ one deforms only $P^-$, using (\ref{21sep10}) we find
 \begin{equation}
 \label{23oct2}
 \delta L^{x-,+}=x \delta T^{-,+} + \delta R^{x-,+}, \qquad \delta L^{\bar x -,+}= \bar x \delta T^{-,+}+\delta R^{\bar x-,+}.
 \end{equation}
 Integrating it over $x^+=0$ and making the Fourier transform, one can express $\delta J^{x-}$ and $\delta J^{\bar x-}$ 
 in terms of  $\delta P^-$. Taking into account definitions (\ref{22sep1}) one can see that (\ref{22sep10}), (\ref{22sep12}) 
 just mean
 that in (\ref{23oct2}) the spin current remains undeformed, $\delta R=0$.
  
With (\ref{22sep10}) imposed, the consistency condition (\ref{22sep9}) simplifies to
   \begin{align}
   \label{22sep11}
  [H_3, J_{n-1}^{z-}] + \dots + [H_{n-1}, J_3^{z-}]+
 \frac{1}{n!}\sum_{\lambda_i}\int d^{3n}q^\perp \delta (\sum_{i=1}^n q^\perp_i){\cal J}^{\{i\}} h_n^{\lambda_1 \dots \lambda_n} 
 \prod_{i=1}^n\Phi^{\lambda_i}(q^\perp_i)=0.
  \end{align}  
 To evaluate  commutators we use
 \begin{equation}
 [F(\Phi),G(\Phi)] = [\Phi^i,\Phi^j] \;\frac{\delta F(\Phi)}{\delta \Phi^i} \frac{\delta G(\Phi)}{\delta \Phi^j}.
 \end{equation}
  A straightforward computation gives
 \begin{align}
 \notag
& [H_n,J_m]+[H_m,J_n] = \frac{1}{(n-1)!}\frac{1}{(m-1)!}\sum_{\lambda_i,\lambda_j} \int d^{3n} q^\perp_i d^{3m}q^\perp_j \delta(\sum_{i=1}^n q^\perp_i)\delta(\sum_{j=n+1}^{n+m} q^\perp_j)
 \\
 & \frac{\delta(q^\perp_{1}+q^\perp_{n+1})\delta^{\lambda_{1}+\lambda_{n+1},0}}{\beta_{1}-\beta_{n+1}}
  \Big( \frac{\partial}{\partial \bar q_{1}}-\frac{\partial}{\partial \bar q_{n+1}} \Big)h_n^{\lambda_{1}\dots \lambda_{n}}
  h_m^{\lambda_{n+1}\dots \lambda_{n+m}}\prod_{i=2}^n\Phi^{\lambda_i}(q^\perp_i)\prod_{j=n+2}^{n+m}\Phi^{\lambda_j}(q^\perp_j).
  \label{22sep13}
 \end{align}
 Note that when $m=n$  one has  $[H_n,J_n]$ only once and thus obtains only  a half of the right hand side of (\ref{22sep13}).
 
\subsection{Identities and notations}
\label{sect2oct1}

So far we were quite explicit with the variables that $h$ depends on, momentum conserving delta-functions, contractions with fields, etc.
To remove unnecessary information that just repeats form line to line, we introduce shortcut notations. Let us illustrate them by the example
of (\ref{22sep13}) which we will write as
\begin{equation}
\label{1oct1}
[H_n,J_m]+[H_m,J_n] = \frac{1}{(n-1)!}\frac{1}{(m-1)!}\frac{1}{\beta_{i_j}-\beta_{j_i}}\Big( \frac{\partial}{\partial \bar q_{i_j}}-\frac{\partial}{\partial \bar q_{j_i}} \Big)
h_n^{\{i\}}(q^\perp_{i_j})h_m^{\{j\}}(q^\perp_{j_i}).
\end{equation}
Here $\{i\}$ refers to the set of indices carried by the variables $h_n^{\{i\}}$ depends on. The set $\{i\}$ has a special
element $i_j$, which is associated with a field, that was removed by the commutator. The same holds for $h_m^{\{j\}}$.
Dependence of $h$ on special momenta will be important, so it is written explicitly.
The momentum conserving delta-functions in our new notations impose
\begin{equation}
\sum_{\{ i\}} q^\perp_i = 0 ,\qquad \sum_{\{ j\}} q^\perp_j = 0 ,\qquad q^\perp_{i_j}+ q^\perp_{j_i}=0.
\end{equation}
In the following they  will be implicit. We will also use 
\begin{equation}
\label{1oct7}
\{i\}_j\equiv \{i\}/i_j, \qquad   \{ i - j\}\equiv \{i\}_j \cup \{j\}_i.
\end{equation}

\begin{figure}
  \centering
  \includegraphics[scale=0.8]{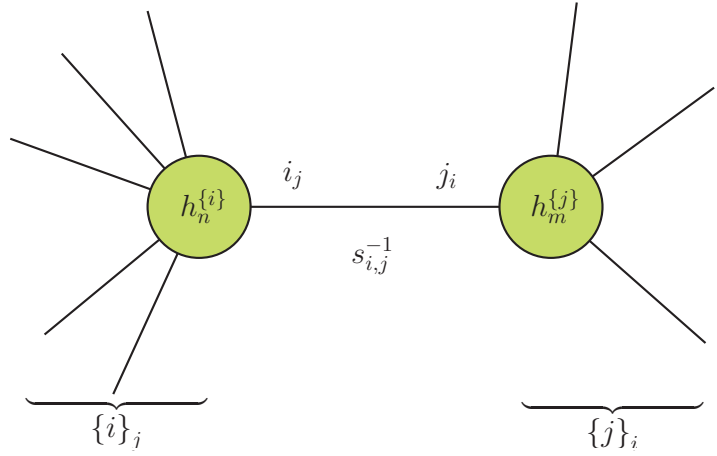}
  \caption{This figure illustrates our conventions on labelling external and internal lines in the case of two
  vertices.}
  \label{fig2vert}
  \end{figure}

Below (\ref{1oct1}) will be related to an exchange involving vertices $h_n$ and $h_m$
with $\{i\}$ and $\{j\}$ labelling fields entering the first and the second vertices respectively. Moreover,
$\{i\}_j$ and $\{j\}_i$ label external legs of the diagram, while $i_j$ and $j_i$ are labels
for the exchanged field, see Figure \ref{fig2vert}.

Next, one has
\begin{align}
 \notag
 (\sum_{\{i\}_j} \vec q_i)^2 &\; \equiv 2 (\sum_{\{i\}_j} q_i)(\sum_{\{i\}_j} \bar q_i)+2(\sum_{\{i\}_j} \beta_i)(\sum_{\{i\}_j} q^-_i)\\
 &\;\approx -2  (-\sum_{\{i\}_j} \beta_i)\Big( h_2 (-\sum_{\{i\}_j} q^\perp_i)+\sum_{\{i\}_j} h_2(q^\perp_i)\Big)=-2\beta_{i_j}{\cal H}^{\{i\}},
 \label{23sep3}
 \end{align}
 where we used $(\vec q)^2\equiv 2q^-q^++2q\bar q$ for the momentum squared.
Then  $s_i$ defined by
 \begin{equation}
 \label{1oct2}
 s_i \equiv - 2 \beta_{i_j}{\cal H}^{\{i\}}
 \end{equation}
 can be interpreted as the Mandelstam variable.
Note that in (\ref{23sep3}) we employed  $q_i^- \approx h_2(q_i^\perp)$ for external particles, which holds only when they
are on-shell. So, (\ref{23sep3}) should be taken just as a motivation for definition (\ref{1oct2}), which is understood off-shell.
We also introduce the symmetric  Mandelstam variable
\begin{equation}
\label{8oct2}
s_{i,j} \equiv \frac{1}{2}(s_i +s_j)= -\frac{1}{2}(\beta_{i_j}-\beta_{j_i})({\cal H}^{\{i\}}-{\cal H}^{\{j\}}).
\end{equation}

Below we will use a slightly modified version of (\ref{1oct3})
\begin{equation}
\label{1oct4}
\big({\cal J}^{\{i\}}-  {\cal H}^{\{i\}}
   \frac{\partial}{\partial \bar q_{i_j}}\big) (\sum_{\{i\}} q^{\perp}_i)\propto (\sum_{\{i\}} q^{\perp}_i)
\end{equation}
as well as 
\begin{equation}
\label{1oct5}
({\cal J}^{\{i\}}+{\cal J}^{\{j\}}) (q_{i_j}^\perp+ q_{j_i}^\perp) \propto (q_{i_j}^\perp+ q_{j_i}^\perp). 
\end{equation}
Finally, we introduce a notation for $q^\perp_{i_j}$ when it is expressed in terms of external
momenta
\begin{equation}
\label{1oct8}
Q^\perp_i \equiv -\sum_{\{i\}_j}q^\perp_i, \qquad Q^\perp_{i,j}=\frac{1}{2}(Q^\perp_i - Q^\perp_j).
\end{equation}

\subsection{Single exchanges}
\label{29oct3}

First we note that 
\begin{align}
\notag
 &\frac{1}{2}({\cal H}^{\{i\}} - {\cal H}^{\{j\}}) \big( \frac{\partial}{\partial \bar q_{i_j}}- \frac{\partial}{\partial \bar q_{j_i}} \big) h_n^{\{i\}}(q^\perp_{i_j})
 h_m^{\{j\}}(q^\perp_{j_i})\\
  \label{25sep8}
& \qquad \qquad \qquad \qquad =\big( {\cal H}^{\{i\}}\frac{\partial}{\partial \bar q_{i_j}} 
 +{\cal H}^{\{j\}}\frac{\partial}{\partial \bar q_{j_i}}\big) 
  h_n^{\{i\}} \big(\tfrac{1}{2}(q^\perp_{i_j}-q^\perp_{j_i})\big)  h_m^{\{j\}} \big(\tfrac{1}{2}(q^\perp_{j_i}-q^\perp_{i_j})\big),
 \end{align}
 where the arguments $q^\perp_{i_j}$ and $q^\perp_{j_i}$ should be replaced before evaluating derivatives as indicated.
Then the right hand side of (\ref{1oct1}) can be written as
\begin{align}
 \notag
& [H^n,J^m]+[H^m, J^n]\\
&\;\;=\frac{1}{(n-1)!}\frac{1}{(m-1)!}\frac{1}{s_{i,j}}\big( -{\cal H}^{\{i\}}\frac{\partial}{\partial \bar q_{i_j}} 
 -{\cal H}^{\{j\}}\frac{\partial}{\partial \bar q_{j_i}}\big) 
  h_n^{\{i\}} \big( \tfrac{1}{2}(q^\perp_{i_j}-q^\perp_{j_i})\big)  h_m^{\{j\}} \big(\tfrac{1}{2}(q^\perp_{j_i}-q^\perp_{i_j})\big),
 \end{align}
 where $s_{i,j}$ was defined in (\ref{8oct2}).
 
 Next we proceed by adding and subtracting ${\cal J}^{\{i\}}+ {\cal J}^{\{j\}}$, so as to produce operators that commute
with the total momentum as in
 (\ref{1oct4})
 \begin{align}
 \notag
 & [H_n,J_m]+[H_m, J_n]\\
 \notag
  & \qquad =\frac{1}{(n-1)!}\frac{1}{(m-1)!}\frac{1}{s_{i,j}}\big({\cal J}^{\{i\}}+ {\cal J}^{\{j\}}- {\cal H}^{\{i\}}\frac{\partial}{\partial \bar q_{i_j}} 
 -{\cal H}^{\{j\}}\frac{\partial}{\partial \bar q_{j_i}}\big) \\
 \notag
  &\qquad\qquad\qquad\qquad\qquad\qquad \qquad\qquad  \qquad \;
  h_n^{\{i\}} \big( \tfrac{1}{2}(q^\perp_{i_j}-q^\perp_{j_i})\big)  h_m^{\{j\}} \big( \tfrac{1}{2}(q^\perp_{j_i}-q^\perp_{i_j})\big)\\
 \label{1oct6}
  &\qquad -\frac{1}{(n-1)!}\frac{1}{(m-1)!}\frac{1}{s_{i,j}} \big({\cal J}^{\{i\}}+ {\cal J}^{\{j\}}\big) h_n^{\{i\}} \big( \tfrac{1}{2}(q^\perp_{i_j}-q^\perp_{j_i})\big)  h_m^{\{j\}} \big( \tfrac{1}{2}(q^\perp_{j_i}-q^\perp_{i_j})\big).
  \end{align}
  The operator that appears in the first term  permits us to use momentum conservation 
  inside $h_m$ and $h_n$, so we can eliminate $q^\perp_{i_j}$ and $q^\perp_{j_i}$ in favour of external momenta. After that $h_n$
  and $h_m$ no longer depend on exchanged momenta explicitly, hence differential operators acting on them can be dropped. As a result,
for the first term in (\ref{1oct6}) we obtain
  \begin{align}
  \notag
  &\frac{1}{(n-1)!}\frac{1}{(m-1)!}\frac{1}{s_{i,j}}\big({\cal J}^{\{i\}}+ {\cal J}^{\{j\}}- {\cal H}^{\{i\}}\frac{\partial}{\partial \bar q_{i_j}} 
 -{\cal H}^{\{j\}}\frac{\partial}{\partial \bar q_{j_i}}\big) \\
 \notag
  &\qquad\qquad\qquad\qquad\qquad\qquad\qquad \qquad \qquad
    h_n^{\{i\}} \big(\tfrac{1}{2}(q^\perp_{i_j}-q^\perp_{j_i})\big)  h_m^{\{j\}} \big( \tfrac{1}{2}(q^\perp_{j_i}-q^\perp_{i_j})\big)\\
  &\qquad\qquad\qquad\qquad\qquad\qquad\quad =\frac{1}{(n-1)!}\frac{1}{(m-1)!}\frac{1}{s_{i,j}}{\cal J}^{\{i-j\}}
  h_n^{\{i\}} ( Q^\perp_{i,j})  h_m^{\{j\}} ( Q^\perp_{j,i}),
  \end{align}
  where $\{i-j\}$ and $Q^\perp_{i,j}$ were defined in (\ref{1oct7}) and (\ref{1oct8}).

  Employing (\ref{1oct5}), the second term reads
  \begin{align}  \notag
  & -\frac{1}{(n-1)!}\frac{1}{(m-1)!}\frac{1}{s_{i,j}} \big({\cal J}^{\{i\}}+ {\cal J}^{\{j\}}\big) h_n^{\{i\}} \big( \tfrac{1}{2}(q^\perp_{i_j}-q^\perp_{j_i})\big)  h_m^{\{j\}} \big( \tfrac{1}{2}(q^\perp_{j_i}-q^\perp_{i_j})\big)\\
  &\qquad\qquad\qquad\qquad\qquad=- \frac{1}{(n-1)!}\frac{1}{(m-1)!}\frac{1}{s_{i,j}}\big({\cal J}^{\{i\}}+{\cal J}^{\{j\}}\big)  h_n^{\{i\}}(q^\perp_{i_j})
  h_m^{\{j\}}(q^\perp_{j_i}).
  \end{align}
  
  Combining both contributions, we find that
  \begin{align}
  \notag
  [H_n,J_m]+[H_m, J_n]
 &= \frac{1}{(n-1)!}\frac{1}{(m-1)!}\frac{1}{s_{i,j}}{\cal J}^{\{i-j\}}
  h_n^{\{i\}} (Q^\perp_{i,j})  h_m^{\{j\}} (Q^\perp_{j,i})\\
   \label{25sep9}
&  - \frac{1}{(n-1)!}\frac{1}{(m-1)!}\frac{1}{s_{i,j}}\big({\cal J}^{\{i\}}+{\cal J}^{\{j\}}\big)  h_n^{\{i\}}(q^\perp_{i_j})h_m^{\{j\}}(q^\perp_{j_i}).
 \end{align}

To interpret this result, first note that in the first term ${\cal J}^{\{ i-j\}}$ acts only on external momenta. Let us denote it just by ${\cal J}$.
In Section \ref{sect1oct1} it will be shown that $s_{i,j}$ commutes with ${\cal J}$. Using these facts one can rewrite the consistency condition
(\ref{22sep11}) as
\begin{align}
\notag
&\sum_{k=0}^{n-2}[H_{n-k},J_{k+2}]={\cal J}\Big( \frac{1}{n!} h_n^{\{i\}} + \sum_{k=1}^{n-3} \frac{1}{(n-k-1)!}\frac{1}{(k+1)!}
  h^{\{i\}}_{n-k} ( Q^\perp_{i,j}) \frac{1}{s_{i,j}} h_{k+2}^{\{j\}} ( Q^\perp_{j,i})\Big)\\
  \label{1oct9}
 & \qquad\qquad\qquad
 -\sum_{k=1}^{n-3} \frac{1}{(n-k-1)!}\frac{1}{(k+1)!}\frac{1}{s_{i,j}}\big({\cal J}^{\{i\}}+{\cal J}^{\{j\}}\big)h_{n-k}^{\{i\}}(q^\perp_{i_j})
 h_{k+2}^{\{j\}}(q^\perp_{j_i})=0.
\end{align}
After multiplying the first line by $n!$ in brackets we recover  the sum of the contact $n$-point diagram and of all exchanges involving
$n$ external particles and
a single propagator. The combinatorial factors 
\begin{equation}
\frac{n!}{(n-k-1)!(k+1)!},
\end{equation}
that appear in front of exchanges count all possible channels that each given exchange can have. As it was noted below (\ref{22sep13}), 
for $n-k = k+2$ the exchange will get an extra factor of $1/2$, which is the standard symmetry factor associated with a symmetry that
interchanges identical vertices.

In other words, up to contributions from diagrams involving more than one propagator, the consistency condition (\ref{1oct9})
looks exactly as the Ward identity for the $n$-point amplitude, where ${\cal J}$ is the operator that verifies gauge invariance. 
At this point  the term Ward identity may sound misleading, because in the light-cone approach
gauge invariance is completely fixed.
We will justify this terminology in Section \ref{sect5oct3}, where we will make a connection between the light-cone
approach and the spinor-helicity  formalism.

\subsection{Double exchanges}
\label{sec29oct2}

 The second term on the right hand side of (\ref{25sep9}) is responsible for contributions of 
 multiple exchanges to the Ward identity. To prevent possible confusions, let us clarify, that by double and
 multiple exchanges we mean tree-level diagrams that involve two or more internal lines.
 
\begin{figure}
  \centering
  \includegraphics[scale=0.8]{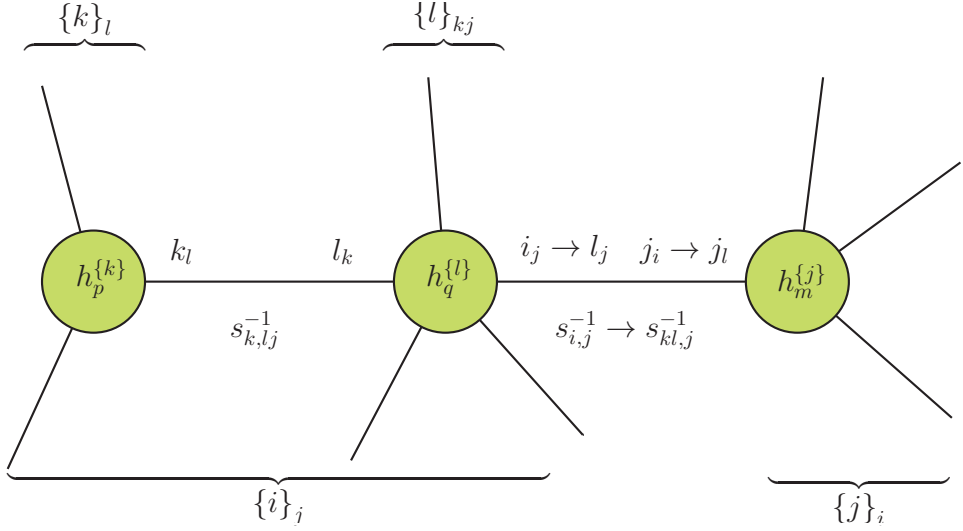}
  \caption{This figure illustrates how after using the consistency condition for $h_n^{\{i\}}$ its external lines previously 
  labelled by $\{i\}_j$ split into two groups labelled by  $\{k\}_l$ and $\{l\}_{kj}$. Accordingly, we rename $i_j\to l_j$, 
  $j_i\to j_l$ and $s_{i,j}\to s_{kl,j}$.}
  \label{fig3vert}
  \end{figure}

 In this Section we will show how to reproduce contributions from diagrams with two exchanges. To this end, let us
 focus on a particular diagram, which consists of two vertices $h^{\{k\}}_p$ and $h^{\{j\}}_m$ connected by a pair
 of exchanges and a vertex  $h^{\{l\}}_q$, 
 see Figure \ref{fig3vert}. The remaining contributions will be omitted, which will be indicated by $\to$
 instead of $=$.
 To have the same number of external fields as in (\ref{25sep9}) we have 
 to demand $p+q-2 = n$. One finds 
 \begin{align}
 \notag
& [H_n,J_m]+[H_m,J_n]  \to - \frac{1}{(n-1)!}\frac{1}{(m-1)!}\frac{1}{s_{i,j}}{\cal J}^{\{i\}}  h_n^{\{i\}}(q^\perp_{i_j})h_m^{\{j\}}(q^\perp_{j_i})\\
\label{2oct1}
 & \to  \frac{1}{(p-1)!}\frac{1}{(q-1)!}\frac{n}{(m-1)!}\frac{1}{s_{kl,j}} \frac{1}{\beta_{k_l}-\beta_{l_k}}
h^{\{j\}}_m(q^\perp_{j_i}) \big(\frac{\partial}{ \partial \bar q_{k_l}}-\frac{\partial}{ \partial \bar q_{l_k}} \big)h^{\{k\}}_p(q^\perp_{k_l})h^{\{l\}}_q(q^\perp_{l_k}).
 \end{align}
 To obtain the last line we used the consistency condition that relates $h_n^{\{i\}}$ to commutators involving the Hamiltonians of lower degrees
 $h^{\{k\}}_p$ and $h^{\{l\}}_q$.
 We also renamed $s_{i,j}\to s_{kl,j}$ to be consistent with the fact that the legs of the diagram, that
 before using the consistency condition for $h_n^{\{i\}}$ were 
 labelled by $\{i\}$, after that belong to  $h^{\{k\}}_p$ and $h^{\{l\}}_q$ and hence are labelled by the sets $\{k\}$ and $\{l\}$.

 We would like to remind the reader, that in the first line of (\ref{2oct1}) there is an implicit delta-function 
 $\delta(q^\perp_{i_j}+q^\perp_{j_i})$, which relates momenta on different sides of the exchange. When we go to 
 the second line, the set $\{ i\}$ is split into $\{k\}$ and $\{l\}$ and
 the special index $i_j$ of the set $\{i\}$ can either belong to $\{k\}$ or to $\{l\}$. Here we keep only the terms, relevant
 to $\{ k-l-j\}$ exchange that is those with  $i_j \in \{l\}$. This produces an extra combinatorial
 factor $(q-1)/n$. As a result
 \begin{align}
 \notag
& [H_n,J_m]+[H_m,J_n]  \\
\label{2oct02}
 &\to  \frac{1}{(p-1)!}\frac{1}{(q-2)!}\frac{1}{(m-1)!}\frac{1}{s_{kl,j}} \frac{1}{\beta_{k_l}-\beta_{l_k}}
 \big(\frac{\partial}{ \partial \bar q_{k_l}}-\frac{\partial}{ \partial \bar q_{l_k}} \big)h^{\{k\}}_p(q^\perp_{k_l})h^{\{l\}}_q(q^\perp_{l_k},q^\perp_{l_j})
 h^{\{j\}}_m(q^\perp_{j_l}),
 \end{align}
 where we also relabelled $i_j$ to $l_j$ and $j_i$ to $j_l$.
 
 Leaving aside the combinatorial factor for a moment, we proceed with the rest analogously to a single exchange case
 \begin{align}
 \notag
 W^{\{k\text{-}l\text{-}j\}}_1&\,\equiv \frac{1}{s_{kl,j}}\frac{1}{\beta_{k_l}-\beta_{l_k}}\big( \frac{\partial}{ \partial \bar q_{k_l}}-
 \frac{\partial}{\partial \bar q_{l_k}}\big) h^{\{k\}}_p(q^\perp_{k_l})h^{\{l\}}_q(q^\perp_{l_k},q^\perp_{l_j})
 h^{\{j\}}_m(q^\perp_{j_l})\\
 \notag
 &\,=-\frac{1}{s_{kl,j}} \frac{1}{s_{k,lj}}({\cal H}^{\{k\}} - {\cal H}^{\{l\}} - {\cal H}^{\{j\}})
\big( \frac{\partial}{ \partial \bar q_{k_l}}-
 \frac{\partial}{\partial \bar q_{l_k}}\big) h^{\{k\}}_p(q^\perp_{k_l})h^{\{l\}}_q(q^\perp_{l_k},q^\perp_{l_j})
 h^{\{j\}}_m(q^\perp_{j_l})
\\
 \notag
 &\,=-\frac{1}{s_{kl,j}} \frac{1}{s_{k,lj}} \big( {\cal H}^{\{k\}} \frac{\partial}{\partial \bar q_{k_l}}+
 ({\cal H}^{\{l\}} + {\cal H}^{\{j\}})\frac{\partial}{\partial \bar q_{l_k}}\big)
  h_p^{\{k\}}\big( \tfrac{1}{2}(q^\perp_{k_l}-q^\perp_{l_k})\big)\\
 \label{25sep12}
 &\,\qquad \qquad\qquad  \qquad\qquad  \quad h_q^{\{l\}}\big(\tfrac{1}{2}(q^\perp_{l_k}-q^\perp_{k_l}),\tfrac{1}{2}(q^\perp_{l_j}-q^\perp_{j_l})\big)
 h_m^{\{j\}}\big( \tfrac{1}{2}(q^\perp_{j_l}-q^\perp_{l_j})\big).
 \end{align}
 In a similar way, from $[H_p,J_{q+m-2}]+[H_{q+m-2},J_p]$ we find another contribution
  \begin{align}
 \notag
 W^{\{k\text{-}l\text{-}j\}}_2&\,\equiv -\frac{1}{s_{kl,j}} \frac{1}{s_{k,lj}} \big( ({\cal H}^{\{k\}} + {\cal H}^{\{l\}})\frac{\partial}{\partial \bar q_{l_j}} +{\cal H}^{\{j\}} \frac{\partial}{\partial \bar q_{j_l}}\big)
  h_p^{\{k\}}\big( \tfrac{1}{2}(q^\perp_{k_l}-q^\perp_{l_k})\big)\\
 \label{2oct2}
 &\,\qquad \qquad\qquad  \qquad\qquad  \quad h_q^{\{l\}}\big(\tfrac{1}{2}(q^\perp_{l_k}-q^\perp_{k_l}),\tfrac{1}{2}(q^\perp_{l_j}-q^\perp_{j_l})\big)
 h_m^{\{j\}}\big( \tfrac{1}{2}(q^\perp_{j_l}-q^\perp_{l_j})\big).
 \end{align}
 
For the sum of two terms we proceed as for the single exchange case in (\ref{1oct6}): we add and subtract ${\cal J}^{\{k\}}+{\cal J}^{\{l\}}+{\cal J}^{\{j\}}$
 producing an operator that allows to use momentum conservation inside the diagram. Namely, 
 \begin{align}
 \notag
 W^{\{k\text{-}l\text{-}j\}}_1&+W^{\{k\text{-}l\text{-}j\}}_2 =\frac{1}{s_{kl,j}} \frac{1}{s_{k,lj}}\big({\cal J}^{\{k\}}+{\cal J}^{\{l\}}+{\cal J}^{\{j\}}\\
\notag
& \qquad \qquad \qquad - {\cal H}^{\{k\}} \frac{\partial}{\partial \bar q_{k_l}}-
 ({\cal H}^{\{l\}} + {\cal H}^{\{j\}})\frac{\partial}{\partial \bar q_{l_k}} -({\cal H}^{\{k\}} + {\cal H}^{\{l\}})\frac{\partial}{\partial \bar q_{l_j}} -{\cal H}^{\{j\}} \frac{\partial}{\partial \bar q_{j_l}}\big)\\
 \notag
& \quad \qquad\qquad\qquad h_p^{\{k\}}\big( \tfrac{1}{2}(q^\perp_{k_l}-q^\perp_{l_k})\big)
 h_q^{\{l\}}\big(\tfrac{1}{2}(q^\perp_{l_k}-q^\perp_{k_l}),\tfrac{1}{2}(q^\perp_{l_j}-q^\perp_{j_l})\big)
 h_m^{\{j\}}\big( \tfrac{1}{2}(q^\perp_{j_l}-q^\perp_{l_j})\big)\\
&  \qquad \qquad  -\frac{1}{s_{kl,j}} \frac{1}{s_{k,lj}}\big({\cal J}^{\{k\}}+{\cal J}^{\{l\}}+{\cal J}^{\{j\}}\big)h^{\{k\}}_p(q^\perp_{k_l})h^{\{l\}}_q(q^\perp_{l_k},q^\perp_{l_j})
 h^{\{j\}}_m(q^\perp_{j_l}).
 \end{align}
Employing (\ref{1oct4}) and (\ref{1oct5}),
 it is not hard to see that the operator in the first term allows to use
 \begin{align}
 \notag
 \sum_{\{k\}} q^\perp_k =0,   \quad  \sum_{\{l\}_j} q^\perp_l+ \sum_{\{j\}_l} q^\perp_j  =0,\\
  \sum_{\{j\}} q^\perp_j =0, \quad \sum_{\{l\}_k} q^\perp_l+ \sum_{\{k\}_l} q^\perp_k  =0.
 \end{align}
 These momentum conservation conditions permit us to eliminate $q^\perp_{k_l}$, $q^\perp_{l_k}$, $q^\perp_{l_j}$ and $q^\perp_{l_j}$
 expressing them in terms of external momenta. After that vertices no longer depend on exchanged momenta, so differential operators acting on
 them can be dropped.
 Thus, we find
 \begin{align}
 \notag
W^{\{k\text{-}l\text{-}j\}}_1+W^{\{k\text{-}l\text{-}j\}}_2
&=\frac{1}{s_{kl,j}} \frac{1}{s_{k,lj}} {\cal J}^{\{ k-l-j\}}
h_p^{\{k\}}( Q^\perp_{k,lj})
 h_q^{\{l\}}(Q^\perp_{lj,k},Q^\perp_{kl,j})
 h_m^{\{j\}}( Q^\perp_{j,kl})\\
 \label{2oct3}
& -\frac{1}{s_{kl,j}} \frac{1}{s_{k,lj}}\big({\cal J}^{\{k\}}+{\cal J}^{\{l\}}+{\cal J}^{\{j\}}\big)h^{\{k\}}_p(q^\perp_{k_l})h^{\{l\}}_q(q^\perp_{l_k},q^\perp_{l_j})
 h^{\{j\}}_m(q^\perp_{j_l}).
\end{align} 
 
 The second line is responsible for contributions of diagrams involving at least three propagators. 
 Reinstating the combinatorial factor from (\ref{2oct02}), for the first line of (\ref{2oct3}) we find
 \begin{align}
 \notag
 &[H_n,J_m]+[H_m,J_n] +[H_p,J_{q+m-2}]+[H_{q+m-2},J_p]\\
 \label{2oct4}
 & \quad \to \frac{1}{(p-1)!}\frac{1}{(q-2)!}\frac{1}{(m-1)!}  {\cal J}
h_p^{\{k\}}( Q^\perp_{k,lj})\frac{1}{s_{kl,j}}
 h_q^{\{l\}}(Q^\perp_{lj,k},Q^\perp_{kl,j})\frac{1}{s_{k,lj}}
 h_m^{\{j\}}( Q^\perp_{j,kl}).
 \end{align}
 This gives one of the double exchange  contributions to the Ward identity involving a contact vertex of degree ${p+q+m-4}$. This contact vertex in our conventions
 goes with a prefactor $1/(p+q+m-4)!$. Normalising the contribution of the contact diagram to unity, we find that the double exchange from  (\ref{2oct4}) 
 appears with a factor 
 \begin{equation}
 \frac{(p+q+m-4)!}{(p-1)!(q-2)!(m-1)!},
 \end{equation}
 which just gives the total number of channels. This is exactly what we should expect from the Feynman rules if we symmetrise them over external fields,
as  in our approach.
 
 \subsection{General case}
 \label{29oct4}
 
 In the previous two Sections we showed that the consistency condition for the non-linear deformation in the light-cone approach (\ref{22sep5})
 can be rewritten as the Ward identity involving contributions from contact diagrams, single exchanges and some extra terms bilinear in vertices
 of degrees lower than that of a contact contribution (\ref{1oct9}). Employing the 
 consistency condition for these extra terms, they were shown to produce contributions of double exchanges plus extra terms, trilinear in
  vertices of yet lower degrees (\ref{2oct3}).
 This procedure should be repeated recursively until it terminates due to ${\cal J}h_3 =0$, see (\ref{22sep11}),
 thus reproducing the Ward identity for the complete amplitude. 
 The equation $[H,\bar J]=0$ is  analogous.
The results of this recursive procedure can be summarised as
\begin{myprop}\label{prop2oct1}
  The light-cone consistency condition can be equivalently rewritten as
  a set of  Ward identities for all $n$-point amplitudes
 \begin{equation}
 \label{2oct5}
 {\cal J}^{\{i\}} {\cal A}_n (q^\perp_i)=0, \qquad \bar {\cal J}^{\{i\}} {\cal A}_n (q^\perp_i)=0,
 \end{equation}
 where 
  \begin{align}
  \notag
{\cal J}^{\{i\}}&\, =\sum_{i=1}^n \Big(-\frac{q_i\bar q_i}{\beta_i}\frac{\partial}{\partial \bar q_i}-q_i \frac{\partial}{\partial \beta_i}+\lambda_i \frac{q_i}{\beta_i}\Big),\\
\bar{\cal J}^{\{i\}}&\, =\sum_{i=1}^n \Big(-\frac{q_i\bar q_i}{\beta_i}\frac{\partial}{\partial  q_i}-\bar q_i \frac{\partial}{\partial \beta_i}+\lambda_i \frac{\bar q_i}{\beta_i}\Big)
 \end{align}
  and ${\cal A}_n (q^\perp_i)$ is the off-shell light-cone amplitude constructed according to the following Feynman rules:
  \begin{itemize}
  \item the propagator,  splitting the external fields into sets labelled by 
  $\{k\}_l$ and $\{l\}_k$,
   is given by
  \begin{equation}
  \label{8nov1}
  \frac{1}{s_{k,l}} = -\frac{2}{(\beta_{k_l}-\beta_{l_k})({\cal H}^{\{k\}}-{\cal H}^{\{l\}})},
  \end{equation}
  where 
  \begin{align}
  \notag
  \beta_{k_l} = -\sum_{\{k\}_l} \beta_{k},& \qquad {\cal H}^{\{k\}} = \sum_{\{k\}_l} h_2(q_k^\perp) + h_2\big(-\sum_{\{k\}_l} q_k^\perp\big)\\
   \beta_{l_k} = -\sum_{\{l\}_k} \beta_{l},& \qquad {\cal H}^{\{l\}} = \sum_{\{l\}_k} h_2(q_l^\perp) + h_2\big(-\sum_{\{l\}_k} q_l^\perp\big).
  \end{align}
  More explanations on our conventions are given in Section \ref{sect2oct1}.
  \item Inside vertices, the momenta of exchanged fields $q_{k_l}^\perp$ and $q_{l_k}^\perp$ should be expressed in terms of momenta of external fields as
  \begin{align}
  \notag
  &q^\perp_{k_l} \to Q_{k,l} = \frac{1}{2}(-\sum_{\{k\}_l} q^\perp_{k}+\sum_{\{l\}_k} q^\perp_l),\\
   &q^\perp_{l_k} \to Q_{l,k} = \frac{1}{2}(-\sum_{\{l\}_k} q^\perp_{l}+\sum_{\{k\}_l} q^\perp_k).
   \label{2oct6}
  \end{align}
  \end{itemize} 
  Alternatively, (\ref{2oct5}) can be written as
\begin{align}
 [A,J_2]=0, \qquad 
\label{9oct2}
  [A,\bar J_2]=0,
\end{align}
where $A$ is made of the amplitude ${\cal A}(q^\perp)$ by contracting it with fields.
\end{myprop}

Let us now make  several comments. First, as it was noted above,  the amplitude ${\cal A}(q^\perp)$  is an off-shell object.
Indeed, in the light-cone approach neither the free Hamiltonian nor interaction terms depend on $q_i^-$. For this simple reason 
${\cal A}(q^\perp)$ defined above makes perfect sense for any $q_i^-$, not only for on-shell values $q_i^- \approx h_2(q_i^\perp)$.
Moreover, ${\cal A}({q^\perp})$ is related unambiguously to the Hamiltonian, so it defines the action.

It is worth to note the unusual form of the propagator (\ref{8nov1}), which we employed to construct ${\cal A}(q^\perp)$.
 It follows from  (\ref{23sep3}) that this propagator coincides with the one
that comes from the covariant Feynman rules  when external particles are on-shell. 
They are, nevertheless, different for off-shell external momental. This difference can be explained by the fact 
that the framework we are dealing with is the light-cone Hamiltonian perturbation theory, which has some peculiar
features, see e.g. \cite{Kogut:1969xa}.

We recall that to derive (\ref{2oct5}) we fixed the freedom of integration by parts in interaction terms by (\ref{22sep10}). 
Of course,  this condition should not be important for consistency. In particular, it should be possible to use
conservation of the total momentum inside ${\cal A}_n(q^\perp)$. The resulting amplitude will no longer be annihilated by ${\cal J}^{\{i\}}$,
but, as it is not hard to see, will produce terms proportional to ${\cal H}^{\{i\}}$. Returning back to the consistency conditions in the original
form (\ref{22sep9}), one can see that such extra terms are harmless, as they can be absorbed by the appropriate redefinition of $j_n$. 
Thus, one can reformulate Proposition \ref{prop2oct1} above in a form, where integration by parts is allowed, but (\ref{2oct5}) holds only up to
terms proportional to ${\cal H}^{\{i\}}$. One can then use the freedom of integration by parts to simplify slightly the Feynman rules presented above.
In particular, it is not necessary to express momenta of internal lines in a symmetric way in terms of external momenta on both sides of a
given propagator as in (\ref{2oct6}).

Note, that  for four external
lines (\ref{2oct5}) was   found in  \cite{Metsaev:1991mt,Metsaev:1991nb}. More precisely, it was
observed that one can solve the light-cone consistency condition at the four-point level by taking the quartic vertex to be minus the sum of
exchanges. Our analysis gives a simple derivation of this fact and extends it to all orders.

Finally, let us point out that even though the form (\ref{2oct5}) of the consistency condition looks much more natural from the
perturbative field theory
perspective, it is sometimes instructive to keep in mind the original form of the consistency condition as well. In particular, it is clear
from (\ref{22sep11}) that once cubic interactions are consistent by themselves, that is $[H_3,J_3]=0$, then by setting
all higher vertices to zero we obtain a consistent theory. At the same time, using the language of (\ref{2oct5}) this translates into the
statement, that once we verified that cubic vertices result into consistent four-point exchanges, then all higher-point
exchanges involving only cubic vertices are also consistent.
The latter statement  is  not so easily seen only from (\ref{2oct5}).
The origin of this non-trivial translation between the two languages is the iterative procedure that we needed to undertake to derive
(\ref{2oct5}) from (\ref{22sep11}). It would be interesting to find more non-trivial examples of this phenomenon.

\section{Solution of the Ward identity}
\label{sect1oct1}

Previous analysis emphasises the central role played by  operators ${\cal J}$ and $\bar{\cal J}$. It was shown that 
consistency of the non-linear deformation 
 implies that these operators annihilate the total amplitude. In this Section we find a general solution
 of these equations.

\subsection{External scalars}
\label{sect5oct2}

To start we find a general solution of (\ref{2oct5}) in the case when all external fields are scalars. The first 
equation explicitly reads
\begin{equation}
\label{3oct2}
{\cal J}^{\{i\}} {\cal A}_n(q_i^\perp) = 0, \qquad {\cal J}^{\{i\}} = \sum_{i=1}^n\left(-\frac{q_i\bar q_i}{\beta_i}\frac{\partial}{\partial \bar q_i}-q_i \frac{\partial}{\partial \beta_i}\right). 
\end{equation}

We remind the reader that the Hamiltonian should satisfy kinematical constraints
(\ref{30sep2})-(\ref{30sep5}). By inspecting the Feynman rules given in the previous Section, it is 
not hard to see that the same kinematical constraints carry over to the total amplitude.

As it was mentioned above, the first two conditions (\ref{30sep2}), (\ref{30sep3}) 
imply that the amplitude is a general function of the following varaibles
\begin{equation}
\label{3oct1}
\bar{\mathbb{P}}_{ij} \equiv \bar q_i\beta_j -\bar q_j \beta_i , \qquad \mathbb{P}_{ij} \equiv q_i\beta_j -q_j \beta_i, \qquad \beta_i.
\end{equation}
This can be seen, for instance, by considering characteristic vector fields associated with differential equations 
(\ref{30sep2}), (\ref{30sep3}). Variables (\ref{3oct1}) provide an overcomplete set of invariants of the associated 
characteristic flows.

When we add an extra differential equation (\ref{3oct2}), clearly, the set of invariants can only be reduced. 
Upon proper dressing with $\beta$ dependence, both $\bar{\mathbb{P}}_{ij}$ and $\mathbb{P}_{ij}$
can be promoted to solutions of (\ref{3oct2})
\begin{equation}
\bar{\mathbb{P}}_{ij} \to \bar\Pi_{ij} \equiv \frac{\bar{\mathbb{P}}_{ij}}{\beta_i\beta_j}: \qquad \qquad 
{\cal J}^{\{i\}} \bar\Pi_{ij}=0,\qquad {\cal J}^{\{i\}} \mathbb{P}_{ij}=0.
\end{equation}
It is not hard to see that there are no solutions of (\ref{3oct2}) that depend on $\beta_i$ alone, so
${\cal A}_n(q_i^\perp)$ is an arbitrary function of
$\bar\Pi_{ij}$ and $\mathbb{P}_{ij}$. It is convenient to phrase this conclusion as
\begin{equation}
\label{3oct4}
{\cal A}_n(q_i^\perp) = \phi(\bar{\mathbb{P}}_{ij},\mathbb{P}_{ij},\beta_i), \qquad -N_{\beta_i} = N_{\langle i|},
\end{equation}
where
\begin{equation}
N_{\langle i|}\equiv \sum_{j=1}^n N_{\mathbb{P}_{ij}}
\end{equation}
counts the total homogeneity degree of all $\mathbb{P}_{ij}$'s with fixed $i$ and  any $j$.

Similarly, to solve
\begin{equation}
\label{3oct3}
\bar{\cal J}^{\{i\}} {\cal A}_n(q_i^\perp) = 0, \qquad \bar{\cal J}^{\{i\}} = \sum_{i=1}^n\left(-\frac{q_i\bar q_i}{\beta_i}\frac{\partial}{\partial  q_i}-\bar q_i \frac{\partial}{\partial \beta_i}\right) 
\end{equation}
we introduce  new variables $\Pi_{ij}$ 
\begin{equation}
{\mathbb{P}}_{ij} \to \Pi_{ij} \equiv \frac{{\mathbb{P}}_{ij}}{\beta_i\beta_j}: \qquad \qquad 
\bar{\cal J}^{\{i\}} \bar{\mathbb{P}}_{ij}=0,\qquad \bar{\cal J}^{\{i\}} \Pi_{ij}=0
\end{equation}
and find that  ${\cal A}_n(q_i^\perp)$ is some function of $\bar{\mathbb{P}}_{ij}$ and $\Pi_{ij}$,
 which is equivalent to 
\begin{equation}
\label{3oct5}
{\cal A}_n(q_i^\perp) = \phi(\bar{\mathbb{P}}_{ij},\mathbb{P}_{ij},\beta_i), \qquad -N_{\beta_i} =  N_{|i]},
\end{equation}
where
\begin{equation}
 N_{|i]}\equiv \sum_{j=1}^n  N_{\bar{\mathbb{P}}_{ij}}
\end{equation}
counts the total homogeneity degree of all $\bar{\mathbb{P}}_{ij}$'s with fixed $i$ and any $j$.

Combining (\ref{3oct4}) and (\ref{3oct5}) we find that 
\begin{equation}
\label{3oct6}
 N_{|i]} = N_{\langle i|}.
\end{equation}
Provided (\ref{3oct6}) is satisfied, we can rewrite the remaining homogeneity constraint as
\begin{equation}
\label{8oct1}
-N_{\beta_i} = \frac{ N_{|i]} +N_{\langle i|}}{2}.
\end{equation}
It  can be solved as
\begin{equation}
\label{3oct7}
{\cal A}_n(q_i^\perp) =  \chi ([ij],\langle ij\rangle),
\end{equation}
where 
\begin{equation}
\label{27sep14}
[ij] \equiv \frac{\sqrt{2}}{\sqrt{\beta_i\beta_j}}\bar{\mathbb{P}}_{ij}, \qquad \langle ij \rangle \equiv -\frac{\sqrt{2}}{\sqrt{\beta_i\beta_j}}\mathbb{P}_{ij}.
\end{equation}
 Here the numerical coefficients have been chosen so as to agree with 
the standard conventions of the spinor-helicity approach  reviewed in  Appendix \ref{ap3oct1}.

In the next Section it will be used that $[ij]$ and $\langle ij\rangle$ can be factorised in terms of helicity spinors
\begin{equation}
\label{5oct1}
| i]_a = 
\frac{2^{\frac{1}{4}}}{\sqrt{\beta_i}}\left(\begin{array}{c}
\bar q_i\\
-\beta_i
\end{array}\right),
\qquad
\langle i |_{\dot b} = \frac{2^{\frac{1}{4}}}{\sqrt{\beta_i}}
\left(\begin{array}{ccc}
 q_i&& -\beta_i
\end{array}\right)
\end{equation}
as
\begin{align}
[ij] &\,=  \varepsilon^{ab} |j]_a |i]_b=
\frac{2^{\frac{1}{4}}}{\sqrt{\beta_j}}
\left(\begin{array}{ccc}
 \bar q_j&& -\beta_j
\end{array}\right)
\left(
\begin{array}{ccc}
0 && 1\\
-1 && 0
\end{array}
\right)
\frac{2^{\frac{1}{4}}}{\sqrt{\beta_i}}\left(\begin{array}{c}
\bar q_i\\
-\beta_i
\end{array}\right),
\\
\langle ij\rangle&\, = \varepsilon^{\dot a\dot b} \langle i|_{\dot a}\langle j|_{\dot b}=
\frac{2^{\frac{1}{4}}}{\sqrt{\beta_i}}
\left(\begin{array}{ccc}
  q_i&& -\beta_i
\end{array}\right)
\left(
\begin{array}{ccc}
0 && 1\\
-1 && 0
\end{array}
\right)
\frac{2^{\frac{1}{4}}}{\sqrt{\beta_j}}\left(\begin{array}{c}
 q_j\\
-\beta_j
\end{array}\right).
\end{align}
 In these terms (\ref{3oct6}) reads
\begin{equation}
\label{3oct8}
N_{| i]} - N_{\langle i |}=0, \qquad N_{| i]} = | i ]\frac{\partial}{\partial | i]}, \qquad N_{\langle i|} = \langle i |\frac{\partial}{\partial \langle i |}.
\end{equation}
Equation (\ref{3oct7}) supplemented with a homogeneity constraint (\ref{3oct8}) gives a general solution of the Ward identities 
(\ref{3oct2}), (\ref{3oct3})  in the case of external scalars.

\subsection{General case}
\label{sect9oct1}

In this Section we prove
\begin{myprop}\label{prop5oct1}
 A general solution of the Ward identities
\begin{align}
\notag
{\cal J}^{\{i\}} {\cal A}_n(q_i^\perp) &\,= 0, \qquad {\cal J}^{\{i\}} = \sum_{i=1}^n\left(-\frac{q_i\bar q_i}{\beta_i}\frac{\partial}{\partial \bar q_i}-q_i \frac{\partial}{\partial \beta_i}+\lambda_i\frac{q_i}{\beta_i}\right),
\\
\label{3oct08}
\bar{\cal J}^{\{i\}} {\cal A}_n(q_i^\perp) &\,= 0, \qquad \bar{\cal J}^{\{i\}} = \sum_{i=1}^n\left(-\frac{q_i\bar q_i}{\beta_i}\frac{\partial}{\partial  q_i}-\bar q_i \frac{\partial}{\partial \beta_i}-\lambda_i\frac{\bar q_i}{\beta_i}\right) 
\end{align}
is given by
\begin{equation}
\label{3oct9}
{\cal A}_n(q_i^\perp) =  \chi ([ij],\langle ij\rangle), 
\end{equation}
where $\chi$ satisfies a homogeneity condition 
\begin{equation}
\label{3oct10}
(-N_{| i]} + N_{\langle i|} +2\lambda_i) \chi ([ij],\langle ij\rangle)=0
\end{equation}
for every $i$. Spinor products are defined in (\ref{27sep14}).
\end{myprop}

To show this, let us first verify that (\ref{3oct9}), (\ref{3oct10}) is indeed a solution. To evaluate how
the differential part of ${\cal J}^{\{i\}}$ acts on $\chi$ we use 
\begin{align}
\notag
-\sum_{i=1}^n\frac{q_i\bar q_i}{\beta_i}\frac{\partial}{\partial \bar q_i}|j] &\,= -\sum_{i=1}^n\frac{q_i}{\beta_i} N_{\bar q_i}
\frac{2^{\frac{1}{4}}}{\sqrt{\beta_j}}\left(\begin{array}{c}
\bar q_j\\
-\beta_j
\end{array}\right)
=-\frac{q_j}{\beta_j} 
\left(\begin{array}{cccc}
1 &&& 0\\
0 && &0
\end{array}\right)|j],
\\
\notag
-\sum_{i=1}^n q_i\frac{\partial}{\partial \beta_i} |j]  &\,= -\sum_{i=1}^n\frac{q_i}{\beta_i} N_{ \beta_i}
\frac{2^{\frac{1}{4}}}{\sqrt{\beta_j}}\left(\begin{array}{c}
\bar q_j\\
-\beta_j
\end{array}\right)=
\frac{1}{2}\frac{q_j}{\beta_j} 
\left(\begin{array}{ccc}
1 && 0\\
0 && -1
\end{array}\right)|j],
\\
-\sum_{i=1}^n q_i\frac{\partial}{\partial \beta_i}  \langle j| &\,=
-\sum_{i=1}^n\frac{q_i}{\beta_i} N_{ \beta_i}
\frac{2^{\frac{1}{4}}}{\sqrt{\beta_j}}\left(\begin{array}{ccc}
  q_j&& -\beta_j
\end{array}\right)
=
\frac{1}{2}\frac{q_j}{\beta_j} 
\langle j|
\left(\begin{array}{ccc}
1 && 0\\
0 && -1
\end{array}\right)
\label{3oct11}
\end{align}
and the Leibniz rule. We find
\begin{align}
\notag
{\cal J}^{\{i\}} \chi ([ij],\langle ij\rangle) =&\, \sum_{i=1}^n \frac{q_i}{\beta_i}\Big( -\frac{1}{2} |i] \frac{\partial}{\partial |i]}
+
\frac{1}{2}\langle i |\left(\begin{array}{ccc}
1 && 0\\
0 && -1
\end{array}\right) \frac{\partial}{\partial \langle i|}
+\lambda_i
\Big)\chi ([ij],\langle ij\rangle) \\
&\qquad \qquad \qquad \qquad \qquad=-\sum_{i=1}^n \frac{q_i}{\beta_i} \langle i |
\left(\begin{array}{cccc}
0 &&& 0\\
0 && &1
\end{array}\right) \frac{\partial}{\partial \langle i|}
\chi ([ij],\langle ij\rangle) ,
\end{align}
where to get to the last line we used the homogeneity condition (\ref{3oct10}). Now we take into account
that in $\chi$ spinors can only appear in the form of spinor contractions
\begin{equation}
{\cal J}^{\{i\}} \chi ([ij],\langle ij\rangle) =-
\sum_{i,k,l=1}^n \frac{q_i}{\beta_i} \langle i |
\left(\begin{array}{cccc}
0 &&& 0\\
0 && &1
\end{array}\right) \frac{\partial\langle kl \rangle}{\partial \langle i|}
\frac{\partial}{\partial\langle kl\rangle}\chi ([ij],\langle ij\rangle).
\end{equation}
Using that 
\begin{align}
\notag
&-
\sum_{i=1}^n \frac{q_i}{\beta_i} \langle i |
\left(\begin{array}{cccc}
0 &&& 0\\
0 && &1
\end{array}\right) \frac{\partial}{\partial \langle i|}\langle kl \rangle =
-\frac{q_k}{\beta_k}\frac{\sqrt{2}}{\sqrt{\beta_k\beta_l}}
\left(\begin{array}{ccc}
  q_k&& -\beta_k
\end{array}\right)
\left(\begin{array}{cccc}
0 &&& 0\\
0 && &1
\end{array}\right)
\left(
\begin{array}{ccc}
0 && 1\\
-1 && 0
\end{array}
\right) \left(\begin{array}{c}
 q_l\\
-\beta_l
\end{array}\right)\\
& \qquad \qquad\qquad\qquad \qquad+
\frac{q_l}{\beta_l}\frac{\sqrt{2}}{\sqrt{\beta_k\beta_l}}
\left(\begin{array}{ccc}
  q_l&& -\beta_l
\end{array}\right)
\left(\begin{array}{cccc}
0 &&& 0\\
0 && &1
\end{array}\right)
\left(
\begin{array}{ccc}
0 && 1\\
-1 && 0
\end{array}
\right) \left(\begin{array}{c}
 q_k\\
-\beta_k
\end{array}\right)=0,
\end{align}
we find 
\begin{equation}
{\cal J}^{\{i\}} \chi ([ij],\langle ij\rangle) =0.
\end{equation}
Analogously one can show that $\chi$ is annihilated by $\bar {\cal J}^{\{i\}}$, so we conclude that (\ref{3oct9}), (\ref{3oct10})
indeed solves (\ref{3oct08}).

Finally, let us assume that there is another solution $\chi'$ of (\ref{3oct08}), which cannot be written in the form  (\ref{3oct9}), (\ref{3oct10}).
Then, as it is not hard to see, $\chi'/\chi$ satisfies the scalar Ward identities (\ref{3oct2}), (\ref{3oct3}) and cannot be written in the
form (\ref{3oct7}), (\ref{3oct8}). This contradicts the results of the previous Section, hence, $\chi'$ does not exist.

In the four-point case the light-cone Ward identity (\ref{3oct08}) was solved previously in
 \cite{Metsaev:1991mt,Metsaev:1991nb,Bengtsson:2016hss}. 
 Our solution extends these result to any number of external points.
 Moreover, 
 unlike previous results,  solution (\ref{3oct9}), (\ref{3oct10}) is not limited to the class of polynomials in transverse momenta.
 This extension is, in fact, important, as the primary meaning of this solution is to give all possible consistent 
 total amplitudes, which are typically non-local due to contributions from exchanges.

\paragraph{Back to  the Mandelstam variables.}
Now we can easily resolve a loose end left from Section 
 \ref{section5oct1} and prove that the Mandelstam variables 
can be pulled through ${\cal J}$ to form exchanges.
 The algebraic part of ${\cal J}$ clearly commutes with $s_{i,j}$, so it remains to prove that $s_{i,j}$
 is annihilated by the differential part of ${\cal J}$. 
 
It is straightforward to compute
\begin{equation}
 (\sum_{\{i\}_j} \vec q_i)^2 = \sum_{k,l\in \{i\}_j} \vec q_k\cdot  \vec q_l \approx \sum_{k,l\in \{i\}_j}
 \big( q_k\bar q_l + \bar q_k q_l - \frac{q_k\bar q_k}{\beta_k}\beta_l- \beta_k\frac{q_l\bar q_l}{\beta_l} \big)
 = \frac{1}{2} \sum_{k,l\in \{i\}_j}{[kl]\langle kl \rangle}.
\end{equation}
Comparing this with   (\ref{23sep3})-(\ref{8oct2}) we find
\begin{equation}
\label{5oct2}
s_i =  \frac{1}{2}\sum_{k,l\in \{i\}_j}{[kl]\langle kl \rangle}, \qquad s_{i,j} =  \frac{1}{4}\sum_{k,l\in \{i\}_j}{[kl]\langle kl \rangle}+ \frac{1}{4}\sum_{k,l\in \{j\}_i}{[kl]\langle kl \rangle}.
\end{equation}
Note that these formulas 
hold off-shell.
They imply that 
 $s_{i,j}$ can  be presented in the form (\ref{3oct7}), (\ref{3oct8}) and consequently commutes 
with ${\cal J}$, as it was argued.  The same applies to $\bar{\cal J}$.

\section{Interpretation}
\label{sect5oct3}

In preceding Sections we first found that the light-cone consistency condition can be 
rewritten in the form of the Ward identity for the amplitude, constructed form the light-cone
Hamiltonian. Then we showed that a general solution of the Ward identity can be 
conveniently presented in terms of spinor products. These spinor products is a basic building block
 of the spinor-helicity approach, which is effectively used for computations of amplitudes
 in theories of massless particles.  In 
particular, the spinor-helicity approach is to large extent responsible
for the existence of extremely compact representations of tree and loop partial amplitudes
in QCD. In this Section, we would like to draw a link between the outcome of our light-cone analysis
and the spinor-helicity approach aiming to interpret the results we found, use the ideas
from the spinor-helicity approach to plot a strategy for construction of massless higher-spin 
interactions and to see whether the light-cone analysis has something new to offer compared to
the spinor-helicity approach.
 We start by briefly reviewing the spinor-helicity approach. Our review is not meant to be 
  self-contained. We refer the
reader to \cite{Dixon:1996wi,Bern:2007dw,Elvang:2013cua} for general reviews and 
to \cite{Benincasa:2007xk,Conde:2016izb} for discussions more focused on the higher-spin case.

In four space-time dimension, any null vector can be represented as 
\begin{equation}
\label{5oct3}
q_{a\dot  b} \equiv q_\mu (\sigma^\mu)_{ a \dot  b}
=\sqrt{2}\left(
\begin{array}{ccc}
q^- && \bar q\\
 q & &-q^+
\end{array}\right)
\approx
\sqrt{2} 
\left(\begin{array}{ccc}
-\frac{q\bar q}{\beta} && \bar q\\
 q && -\beta
\end{array}\right)= - | q]_a \langle q|_{\dot b},
\end{equation}
where  $| q]_a$ and $\langle q|_{\dot b}$ are simply  on-shell Weyl spinors with momentum
$q$ 
\begin{equation}
\label{9oct1}
| q]_a = 
\frac{2^{\frac{1}{4}}}{\sqrt{\beta}}\left(\begin{array}{c}
\bar q\\
-\beta
\end{array}\right),
\qquad
\langle q |_{\dot b} = \frac{2^{\frac{1}{4}}}{\sqrt{\beta}}
\left(\begin{array}{ccc}
 q&& -\beta
\end{array}\right).
\end{equation}
In the spinor-helicity approach representation (\ref{5oct3}) is used to define momenta of massless
particles of any spin. We will  use the standard notation $|i] \equiv | q_i]$ and $\langle i | \equiv \langle q_i |$.
More conventions are given in Appendix \ref{ap3oct1}.

To encode polarisations
 one uses  auxiliary massless vectors called reference 
momenta. For example, the polarisation vector $\epsilon^\mu_1(q;k)$ of a helicity one boson of momentum $q$ is
defined as
\begin{equation}
\label{5oct4}
\epsilon_1^\mu(q;k)=-\frac{\langle k| \gamma^\mu | q]}{\sqrt{2}\langle k q\rangle},
\end{equation}
where $k$ is the reference momentum. It is easy to see that the polarisation vector defined above  is transversal to $q$, that is $q_\mu \epsilon_1^\mu(q;k)=0$, as required by the Ward identity.
The arbitrariness in the choice of the reference momentum is just a manifestation of gauge redundancy.
For each external field one is free to choose a reference momentum independently. However, this choice should
be consistent for all diagrams relevant to the process. Then, as a consequence of gauge invariance, auxiliary
reference vectors drop out from the final answer. In other words, a consistent amplitude should be a 
function of spinor products $[ij]$ and $\langle ij \rangle$ only.

It is clear from (\ref{5oct3}) that  $q_\mu$ is invariant with respect to the following scaling
transformations
\begin{equation}
\label{23oct3}
 \langle q| \to t\langle q|,
\qquad  | q] \to t^{-1} | q].
\end{equation}
For $|\lambda|=1$ this generates the action of the Wigner little group on the helicity spinors.
 If we compute amplitudes using the
 Feynman rules, then we can see that the only way this scaling contributes to the amplitude is 
 through polarisation vectors.  For a helicity one boson the polarisation vector (\ref{5oct4})  scales as
 \begin{equation}
 \epsilon_1^\mu(q;k) \to t^{-2}  \epsilon_1^\mu(q;k).
 \end{equation}
 More generally, the Wigner little group acts on the helicity-$\lambda$ polarisation vector as
  \begin{equation}
  \label{9oct3}
 \epsilon_1^\mu(q;k) \to t^{-2\lambda}  \epsilon_1^\mu(q;k).
 \end{equation} 
 Each external field can be subjected to   (\ref{23oct3}) independently.
 For an amplitude expressed in terms of spinor contractions to reproduce this scaling
 behaviour
 it should satisfy 
\begin{equation}
\label{9oct4}
( -N_{| i]}  + N_{\langle i |} +2\lambda_i ) {\cal A}_n([jk],\langle jk\rangle ) = 0.
\end{equation}

We find that despite a slightly different motivation, the spinor-helicity approach produces the 
same constraints that we found previously from the Poincare algebra closure in the light-cone 
deformation procedure. As we just reviewed, in the spinor-helicity approach the only two constraints are  gauge
invariance, which requires the amplitude to be expressed in terms of spinor products, and 
invariance  with respect to the action of the Wigner little group, which fixes the homogeneity
degrees of spinors. In the light-cone approach the only constraint is invariance with respect to
the fully non-linear action of the Poincare algebra, which
according to Proposition \ref{prop2oct1}  can be replaced by invariance of the amplitude with respect to the linear action of 
the algebra generators. Solving these constraints, we indeed find that 
the light-cone amplitude can be expressed  in terms of spinor products only, which 
through the spinor-helicity approach relates
(\ref{3oct08}) to gauge invariance and, thus, justifies the term the Ward identity
 that we used\footnote{Note that in the spinor-helicity terms the light-cone gauge (\ref{29sep4}) can  be viewed as 
 a particular choice of the reference vector  along $x^-$ direction.}.
Moreover, (\ref{3oct08}) fixes the homogeneity degrees of spinors, which from the spinor-helicity
perspective is related to the Wigner little group invariance. This relation can be easily seen 
from the light-cone approach itself. Namely, by acting 
 along the lines of Section \ref{sect9oct1},
one can show that $[{\cal A}_n,J_2^{x\bar x}]=0$ implies (\ref{3oct10}).
 
So far we could see that  on-shell amplitudes from the spinor-helicity approach 
appear to be very similar to  light-cone amplitudes ${\cal A}_n(q^\perp)$, which 
we introduced above.
 However, it is important to remember that ${\cal A}_n(q^\perp)$
in the light-cone approach
 is well defined for off-shell momenta.
 In particular, ${\cal A}_n(q^\perp)$ can be used to define the light-cone Hamiltonian and then the
 action via the Legendre transform. This difference originates from the way one defines  
 spinors $|i]$ and $\langle i|$ in the two approaches. In the spinor-helicity approach spinors $|i]$ and $\langle i|$
 are only defined for null momenta, that is for $q^-\approx h_2(q^\perp)$, see (\ref{5oct3}), (\ref{9oct1}).
 In the light-cone approach one defines spinors by the very same formula (\ref{9oct1}), but the momentum
 is not required to be massless and $q^-$ can be arbitrary. Clearly, factorisation formula (\ref{5oct3})
 does not work off-shell, but in the light-cone approach it is never used\footnote{This off-shell continuation appeared in \cite{Bardeen:1995gk,Cangemi:1996rx}.
Later it
was used in \cite{Cachazo:2004kj} to extend  Yang-Mills MHV amplitudes off-shell, which were then treated as vertices in the 
action, see also \cite{Gorsky:2005sf,Mansfield:2005yd}.}. 
It is quite remarkable that this rather trivial off-shell extension of spinor products and, hence, of the amplitude
 turns out to be consistent 
without any further constraints.  This can be regarded as a 
 simple consequence of $q^-$-independence of the Feynman rules presented in Proposition \ref{prop2oct1}.

An immediate benefit from the off-shell extension of helicity spinors is that they can be used
for momenta on internal lines. So, in the light-cone approach exchanges can also be written in terms
of spinor products. More details on how they are constructed can be found in Proposition \ref{prop2oct1}.
 Generically, when internal momenta are expressed in terms of momenta on external lines,
 exchanges cannot be written in terms of spinor products any more. 
 This implies that they do not satisfy the Ward identity. Consequently, individual
contact interactions, in general, violate the Ward identity either. 

This suggests to 
reconsider  the light-cone deformation procedure  and focus on
seeking the amplitude  instead of the  Hamiltonian. The benefit
of this strategy is concisely summarised by (\ref{9oct2}): the amplitude satisfies a simple
linear differential equation, while the equation for the Hamiltonian is
quadratic in deformations. Having found a general solution for the amplitude we,
thus,  found a general solution of the light-cone deformation procedure.
Note, however, that these solutions generically are associated with non-local Hamiltonians.
The problem of finding consistent amplitudes that result in local Hamiltonians deserves a separate 
thorough analysis. Let us, nevertheless,  make few comments on this point.
 
 \paragraph{Locality.}
 Typically, in the light-cone approach one defines locality as a requirement that the Hamiltonian is
 polynomial in transverse momenta or, equivalently, in $\bar{\mathbb{P}}_{ij}$ and ${\mathbb{P}}_{kl}$. 
 The same Hamiltonian
 can brought to many different forms using  momentum conservation. Clearly, the 
 Hamiltonian is  local if there exist at least one of its  forms where it is
 polynomial in transverse momenta. Due to the possibility to use momentum conservation locality
 may not be manifest. For example, the antiholomorphic part of the Yang-Mills cubic vertex can be written as
 \begin{equation}
 \label{10dec1}
 h_{3}= \frac{[12]^3}{[23][31]}=\sqrt{2}\frac{\bar{\mathbb{P}}^3_{12}\beta_3}{\bar{\mathbb{P}}_{23}\bar{\mathbb{P}}_{31}\beta_1\beta_2}.
 \end{equation}
 This vertex is superficially non-local as it contains powers of transverse momenta in the denominator. However,
 it is easy to see that momentum conservation implies 
 \begin{equation}
 \label{10dec2}
 \bar{\mathbb{P}}_{12}=\bar{\mathbb{P}}_{23}=\bar{\mathbb{P}}_{31},
 \end{equation}
 hence, non-locality of (\ref{10dec1}) is spurious. 
 This example illustrates that to make locality manifest it may be required
 to break the spinor-helicity representation. At the cubic level  this subtlety does not
 result in any difficulties, because modulo momentum conservation there is only one possible 
 Lorentz-invariant variable that depends on the antiholomorphic momentum (\ref{10dec2}) and, similarly, only
 one that depends on the holomorphic one. It would be interesting to clarify how this phenomenon extends
 to higher-point amplitudes.
  
 Locality of the Hamiltonian is naturally translated into the language of amplitudes using the framework of
 on-shell methods  \cite{Britto:2004ap,Britto:2005fq}. Namely, it is required that amplitudes have no
 other singularities than those produced by exchanges. In many cases one can also justify that amplitudes
 vanish in certain directions at complex infinity. 
This allows to reconstruct them unambiguously  from their singularities.
   In this respect we would
like to clarify that the light-cone deformation procedure alone does not impose any constraints on the behaviour of 
amplitudes at infinity\footnote{See \cite{Benincasa:2011kn,Benincasa:2011pg,McGady:2013sga} for extensions of 
the on-shell methods, which do not require constraints
on amplitudes at infinite momenta.}. Let us also note that the solution  (\ref{3oct9}), (\ref{3oct10}) 
is not limited to  polynomials in transverse momenta, 
so it is  applicable to amplitudes, which are typically non-local
due to contributions from exchanges.

 It is worth to remark separately that in higher-spin theories
 imposing locality  as described above, most likely, would rule out any interactions at all.
  It is then suggestive to replace locality in a strict sense by 
 a milder   requirement that coefficients of higher derivative terms decrease fast enough, 
 so that the amplitudes associated with contact interactions do not contain singularities. 
 This weaker version of locality was discussed at length in \cite{Bekaert:2015tva}.
 Let us also note that it is this weaker version of locality, that is effectively implemented by the
 on-shell methods, as they just require contact interactions to be free of singularities.
 
Finally, we remark that contrary to the way one usually defines locality  for the light-cone deformation procedure,
 the on-shell methods require locality of interactions only on-shell. It would be interesting 
 to clarify whether this difference can play any role.

\section{Conclusion}
\label{sec7oct2}

This paper contains two main results. First,
 we show that the light-cone consistency conditions can be equivalently
rephrased as a set of  Ward identities for the light-cone off-shell amplitudes. Then we give a general solution 
to these Ward identities. This solution acquires an extremely simple form when written using  the spinor-helicity
language. More precisely, the general solution is just any function of spinor products that satisfies a well-known
constraint 
 relating homogeneity degrees of spinors with helicities of external fields.
 These results are summarised in Proposition \ref{prop2oct1} 
and Proposition \ref{prop5oct1}.

 Our primary goal is to employ the light-cone analysis to construct interactions of massless higher-spin fields. In this
respect, our results provide a general solution to this problem in the case when locality of interactions is not required.
Of course, this way of solving the consistent interaction problem is to large extent trivial, see, e.g., \cite{Barnich:1993vg}.
Nevertheless, this gives a good starting point to address the problem of local interactions.

To construct local interactions, it is suggestive to proceed in the spirit of the on-shell methods
\cite{Britto:2004ap,Britto:2005fq}, that is
  by reconstructing  amplitudes from singularities associated with exchanges. 
  In other words, it seems more reasonable to seek not the individual vertices, but the total amplitude.
   Indeed, the total amplitude satisfies
  the Ward identities, which have been solved in the present paper in complete generality. 
   At the same time, constraints on individual contact
  interactions are much more complicated and from particular examples we know that individual vertices can be quite cumbersome
   \cite{Ponomarev:2016lrm}.
  More generally, amplitudes, being physically observable quantities, are much more constrained than 
   individual vertices which are, moreover,  prone to ambiguities of a gauge choice, field redefinitions, different sets of auxiliary fields etc.
   One should not expect a simple form of individual vertices unless these ambiguities are fixed wisely. 
   On the contrary, as a result of strong constraints put on them,  
   amplitudes admit a concise spinor-helicity representation.

    Having written the light-cone consistency condition in the spinor-helicity form, we are able to clarify that the light-cone
   analysis does not impose any constraints on the behaviour of the amplitude at infinity. Hence, the no-go conclusions
   based on BCFW \cite{Benincasa:2007xk,Fotopoulos:2010ay,Benincasa:2011pg,McGady:2013sga,Ponomarev:2016jqk,Bengtsson:2016alt},
   in principle,  
    can be circumvented, see also \cite{Bengtsson:2016hss}.
     Moreover, the explicit analysis  of the quartic self-interaction sector  \cite{Ponomarev:2016lrm}
   shows that the relevant consistent interaction does exist.

It is worth to emphasise that the light-cone approach leads to a natural off-shell continuation of the spinor-helicity
representation, which is usually defined on-shell.
 Once continued off-shell the light-cone amplitudes unambiguously define the action
of the theory. 
It would be interesting to see whether the light-cone  off-shell continuation  can be used to promote on-shell 
results, such as colour-kinematics duality \cite{Bern:2008qj,Bern:2010ue},  to the off-shell level.
Other interesting directions include extensions of the spinor-helicity approach to massive particles\footnote{One natural way
to do that is to represent massive momenta by pairs of massless ones \cite{Conde:2016vxs,Conde:2016izb}.},
 as well as to AdS.

\section*{Acknowledgements}
 
 I am grateful to E. Skvortsov for many stimulating discussions. I would also like to thank
 R. Metsaev and A. Tseytlin for useful comments on the draft. I am grateful to
  A. Ochirov for explanations on the spinor-helicity approach and to W. Siegel for 
  helpful correspondence.
I acknowledge a kind hospitality at the program ``Higher Spin Theory and Duality" MIAPP, Munich (May 2-27, 2016) organized by the Munich Institute for Astro- and Particle Physics (MIAPP). This work was supported by the ERC Advanced grant No.290456.

\appendix
\section{Notations}
\label{ap3oct1}

Here we collected various notations used throughout the paper.

\paragraph{Light-cone coordinates.}
We work with the 4d Minkowski space endowed with the  mostly plus 
metric 
\begin{equation}
ds^2 = -(dx^0)^2+ (dx^1)^2+(dx^2)^2+(dx^3)^2.
\end{equation}
 In the light-cone coordinates
\begin{align}
\notag
x^+ &\,= \frac{1}{\sqrt{2}}(x^3+x^0), & x^-&\, = \frac{1}{\sqrt{2}}(x^3-x^0),\\
\label{29sep1}
x &\,=\frac{1}{\sqrt{2}}(x^1-ix^2), & \bar x &\,= \frac{1}{\sqrt{2}}(x^1+ix^2),
\end{align}
it becomes
\begin{equation}
ds^2 = 2dx^+ dx^- + 2 dx d\bar x.
\end{equation}
Accordingly, we denote
 \begin{align}
 \notag
\partial^-  &\,= \frac{1}{\sqrt{2}}(\partial^3-\partial^0), &  \partial^+ &\, = \frac{1}{\sqrt{2}}(\partial^3 + \partial^0),\\
\bar\partial &\,= \frac{1}{\sqrt{2}}(\partial^1-i\partial^2), &  \partial &\, =\frac{1}{\sqrt{2}}(\partial^1 + i \partial^2),
\end{align}
 which implies
\begin{equation}
\partial^+ x^- = \partial^- x^+ = \bar\partial x = \partial\bar x = 1.
\end{equation}
In the light-cone approach $x^+$ is taken to be the time variable and $\partial^-$ is the time derivative.
Moreover, one assumes that $\partial^+$ is non-zero and can always be inverted.

\paragraph{Spinor-helicity.}
For spinor-helicity  conventions we follow \cite{Elvang:2013cua}. We choose 
the Pauli matrices as
\begin{equation}
\sigma^0 = 
\left(\begin{array}{cccc}
1 &&& 0\\
0 && &1
\end{array}\right), \quad \sigma^1 = 
\left(\begin{array}{cccc}
0 & && 1\\
1& && 0
\end{array}\right), 
\quad 
 \sigma^2 = 
\left(\begin{array}{ccc}
0 && -i\\
i&& 0
\end{array}\right), \quad
 \sigma^3 = 
\left(\begin{array}{ccc}
1 && 0\\
0&& -1
\end{array}\right).
\end{equation}
Then
\begin{equation}
\label{27sep10}
q_{a\dot  b} \equiv q_\mu (\sigma^\mu)_{ a \dot  b}
=\sqrt{2}\left(
\begin{array}{ccc}
q^- && \bar q\\
 q & &-q^+
\end{array}\right)
\approx
\sqrt{2} 
\left(\begin{array}{ccc}
-\frac{q\bar q}{\beta} && \bar q\\
 q && -\beta
\end{array}\right)= - | q]_a \langle q|_{\dot b},
\end{equation}
where
\begin{equation}
\label{27sep12}
| q]_a = 
\frac{2^{\frac{1}{4}}}{\sqrt{\beta}}\left(\begin{array}{c}
\bar q\\
-\beta
\end{array}\right),
\qquad
\langle q |_{\dot b} = \frac{2^{\frac{1}{4}}}{\sqrt{\beta}}
\left(\begin{array}{ccc}
 q&& -\beta
\end{array}\right).
\end{equation}
In these terms
\begin{equation}
\label{27sep13}
[pq] = [ p |^a |q]_a = \varepsilon^{ab} |q]_a |p]_b, \qquad \langle p q\rangle = \langle p|_{\dot a} |q\rangle^{\dot a} = \varepsilon^{\dot a\dot b} \langle p|_{\dot a}\langle q|_{\dot b},
\end{equation}
where
\begin{equation}
\varepsilon^{ab} = \varepsilon^{\dot a\dot b} = 
\left(
\begin{array}{ccc}
0 && 1\\
-1 && 0
\end{array}
\right) = - \varepsilon_{ab}= -\varepsilon_{\dot a\dot b}.
\end{equation}
Rewriting spinor contractions as matrix products we find
\begin{align}
[ij]&\, = 
\frac{2^{\frac{1}{4}}}{\sqrt{\beta_j}}
\left(\begin{array}{ccc}
 \bar q_j&& -\beta_j
\end{array}\right)
\left(
\begin{array}{ccc}
0 && 1\\
-1 && 0
\end{array}
\right)
\frac{2^{\frac{1}{4}}}{\sqrt{\beta_i}}\left(\begin{array}{c}
\bar q_i\\
-\beta_i
\end{array}\right)=\frac{\sqrt{2}}{\sqrt{\beta_i\beta_j}}\bar{\mathbb{P}}_{ij},
\\
\langle ij\rangle &\,= 
\frac{2^{\frac{1}{4}}}{\sqrt{\beta_i}}
\left(\begin{array}{ccc}
  q_i&& -\beta_i
\end{array}\right)
\left(
\begin{array}{ccc}
0 && 1\\
-1 && 0
\end{array}
\right)
\frac{2^{\frac{1}{4}}}{\sqrt{\beta_j}}\left(\begin{array}{c}
 q_j\\
-\beta_j
\end{array}\right)=-\frac{\sqrt{2}}{\sqrt{\beta_i\beta_j}}{\mathbb{P}}_{ij}.
\end{align}

\bibliography{lcsh}

\providecommand{\href}[2]{#2}\begingroup\raggedright\begin{thebibliography}{10}

\bibitem{Wigner:1939cj}
E.~P. Wigner, ``{On Unitary Representations of the Inhomogeneous Lorentz
  Group},'' {\em Annals Math.} {\bf 40} (1939) 149--204.
[Reprint: Nucl. Phys. Proc. Suppl.6,9(1989)].

\bibitem{Bekaert:2006py}
X.~Bekaert and N.~Boulanger, ``{The Unitary representations of the Poincare
  group in any spacetime dimension},'' in {\em {2nd Modave Summer School in
  Theoretical Physics Modave, Belgium, August 6-12, 2006}}.
\newblock 2006.
\newblock
\href{http://arXiv.org/abs/hep-th/0611263}{{\tt hep-th/0611263}}.
\newblock

\bibitem{Aragone:1979hx}
C.~Aragone and S.~Deser, ``{Consistency Problems of Hypergravity},'' {\em Phys.
  Lett.} {\bf B86} (1979)
161--163.

\bibitem{Bekaert:2010hp}
X.~Bekaert, N.~Boulanger, and S.~Leclercq, ``{Strong obstruction of the
  Berends-Burgers-van Dam spin-3 vertex},'' {\em J. Phys.} {\bf A43} (2010)
  185401,
\href{http://arXiv.org/abs/1002.0289}{{\tt 1002.0289}}.

\bibitem{Joung:2013nma}
E.~Joung and M.~Taronna, ``{Cubic-interaction-induced deformations of
  higher-spin symmetries},'' {\em JHEP} {\bf 03} (2014) 103,
\href{http://arXiv.org/abs/1311.0242}{{\tt 1311.0242}}.

\bibitem{Weinberg:1964ew}
S.~Weinberg, ``{Photons and Gravitons in s Matrix Theory: Derivation of Charge
  Conservation and Equality of Gravitational and Inertial Mass},'' {\em Phys.
  Rev.} {\bf 135} (1964)
B1049--B1056.

\bibitem{Coleman:1967ad}
S.~R. Coleman and J.~Mandula, ``{All Possible Symmetries of the S Matrix},''
  {\em Phys. Rev.} {\bf 159} (1967)
1251--1256.

\bibitem{Bekaert:2010hw}
X.~Bekaert, N.~Boulanger, and P.~Sundell, ``{How higher-spin gravity surpasses
  the spin two barrier: no-go theorems versus yes-go examples},'' {\em Rev.
  Mod. Phys.} {\bf 84} (2012) 987--1009,
\href{http://arXiv.org/abs/1007.0435}{{\tt 1007.0435}}.

\bibitem{Bengtsson:1983pd}
A.~K.~H. Bengtsson, I.~Bengtsson, and L.~Brink, ``{Cubic Interaction Terms for
  Arbitrary Spin},'' {\em Nucl. Phys.} {\bf B227} (1983)
31--40.

\bibitem{Bengtsson:1983pg}
A.~K.~H. Bengtsson, I.~Bengtsson, and L.~Brink, ``{Cubic Interaction Terms for
  Arbitrarily Extended Supermultiplets},'' {\em Nucl. Phys.} {\bf B227} (1983)
41--49.

\bibitem{Bengtsson:1986kh}
A.~K.~H. Bengtsson, I.~Bengtsson, and N.~Linden, ``{Interacting Higher Spin
  Gauge Fields on the Light Front},'' {\em Class. Quant. Grav.} {\bf 4} (1987)
1333.

\bibitem{Conde:2016izb}
E.~Conde, E.~Joung, and K.~Mkrtchyan, ``{Spinor-Helicity Three-Point Amplitudes
  from Local Cubic Interactions},'' {\em JHEP} {\bf 08} (2016) 040,
\href{http://arXiv.org/abs/1605.07402}{{\tt 1605.07402}}.

\bibitem{Sleight:2016xqq}
C.~Sleight and M.~Taronna, ``{Higher-Spin Algebras, Holography and Flat
  Space},''
\href{http://arXiv.org/abs/1609.00991}{{\tt 1609.00991}}.

\bibitem{Metsaev:1991mt}
R.~R. Metsaev, ``{Poincare invariant dynamics of massless higher spins: Fourth
  order analysis on mass shell},'' {\em Mod. Phys. Lett.} {\bf A6} (1991)
359--367.

\bibitem{Metsaev:1991nb}
R.~R. Metsaev, ``{S matrix approach to massless higher spins theory. 2: The
  Case of internal symmetry},'' {\em Mod. Phys. Lett.} {\bf A6} (1991)
2411--2421.

\bibitem{Metsaev:1993ap}
R.~R. Metsaev, ``{Generating function for cubic interaction vertices of higher
  spin fields in any dimension},'' {\em Mod. Phys. Lett.} {\bf A8} (1993)
2413--2426.

\bibitem{Bengtsson:2014qza}
A.~K.~H. Bengtsson, ``{A Riccati type PDE for light-front higher helicity
  vertices},'' {\em JHEP} {\bf 09} (2014) 105,
\href{http://arXiv.org/abs/1403.7345}{{\tt 1403.7345}}.

\bibitem{Boulanger:2006gr}
N.~Boulanger and S.~Leclercq, ``{Consistent couplings between spin-2 and spin-3
  massless fields},'' {\em JHEP} {\bf 11} (2006) 034,
\href{http://arXiv.org/abs/hep-th/0609221}{{\tt hep-th/0609221}}.

\bibitem{Ponomarev:2016lrm}
D.~Ponomarev and E.~D. Skvortsov, ``{Light-Front Higher-Spin Theories in Flat
  Space},''
\href{http://arXiv.org/abs/1609.04655}{{\tt 1609.04655}}.

\bibitem{Ananth:2012un}
S.~Ananth, ``{Spinor helicity structures in higher spin theories},'' {\em JHEP}
  {\bf 11} (2012) 089,
\href{http://arXiv.org/abs/1209.4960}{{\tt 1209.4960}}.

\bibitem{Akshay:2014qea}
Y.~S. Akshay and S.~Ananth, ``{Factorization of cubic vertices involving three
  different higher spin fields},'' {\em Nucl. Phys.} {\bf B887} (2014)
  168--174,
\href{http://arXiv.org/abs/1404.2448}{{\tt 1404.2448}}.

\bibitem{Benincasa:2007xk}
P.~Benincasa and F.~Cachazo, ``{Consistency Conditions on the S-Matrix of
  Massless Particles},''
\href{http://arXiv.org/abs/0705.4305}{{\tt 0705.4305}}.

\bibitem{Bengtsson:2016jfk}
A.~K.~H. Bengtsson, ``{Notes on Cubic and Quartic Light-Front Kinematics},''
\href{http://arXiv.org/abs/1604.01974}{{\tt 1604.01974}}.

\bibitem{Bengtsson:2016alt}
A.~K.~H. Bengtsson, ``{Quartic amplitudes for Minkowski higher spin},'' in {\em
  {International Workshop on Higher Spin Gauge Theories Singapore, Singapore,
  November 4-6, 2015}}.
\newblock 2016.
\newblock
\href{http://arXiv.org/abs/1605.02608}{{\tt 1605.02608}}.
\newblock

\bibitem{Chalmers:1998jb}
G.~Chalmers and W.~Siegel, ``{Simplifying algebra in Feynman graphs. Part 2.
  Spinor helicity from the space-cone},'' {\em Phys. Rev.} {\bf D59} (1999)
  045013,
\href{http://arXiv.org/abs/hep-ph/9801220}{{\tt hep-ph/9801220}}.

\bibitem{Heinzl:2000ht}
T.~Heinzl, ``{Light cone quantization: Foundations and applications},'' {\em
  Lect. Notes Phys.} {\bf 572} (2001) 55--142,
\href{http://arXiv.org/abs/hep-th/0008096}{{\tt hep-th/0008096}}.

\bibitem{Dirac:1949cp}
P.~A.~M. Dirac, ``{Forms of Relativistic Dynamics},'' {\em Rev. Mod. Phys.}
  {\bf 21} (1949)
392--399.

\bibitem{Kogut:1969xa}
J.~B. Kogut and D.~E. Soper, ``{Quantum Electrodynamics in the Infinite
  Momentum Frame},'' {\em Phys. Rev.} {\bf D1} (1970)
2901--2913.

\bibitem{Bengtsson:2016hss}
A.~K.~H. Bengtsson, ``{Investigations into Light-front Quartic Interactions for
  Massless Fields (I): Non-constructibility of Higher Spin Quartic
  Amplitudes},''
\href{http://arXiv.org/abs/1607.06659}{{\tt 1607.06659}}.

\bibitem{Dixon:1996wi}
L.~J. Dixon, ``{Calculating scattering amplitudes efficiently},'' in {\em {QCD
  and beyond. Proceedings, Theoretical Advanced Study Institute in Elementary
  Particle Physics, TASI-95, Boulder, USA, June 4-30, 1995}}, pp.~539--584.
\newblock 1996.
\newblock
\href{http://arXiv.org/abs/hep-ph/9601359}{{\tt hep-ph/9601359}}.
\newblock

\bibitem{Bern:2007dw}
Z.~Bern, L.~J. Dixon, and D.~A. Kosower, ``{On-Shell Methods in Perturbative
  QCD},'' {\em Annals Phys.} {\bf 322} (2007) 1587--1634,
\href{http://arXiv.org/abs/0704.2798}{{\tt 0704.2798}}.

\bibitem{Elvang:2013cua}
H.~Elvang and Y.-t. Huang, ``{Scattering Amplitudes},''
\href{http://arXiv.org/abs/1308.1697}{{\tt 1308.1697}}.

\bibitem{Bardeen:1995gk}
W.~A. Bardeen, ``{Selfdual Yang-Mills theory, integrability and multiparton
  amplitudes},'' {\em Prog. Theor. Phys. Suppl.} {\bf 123} (1996)
1--8.

\bibitem{Cangemi:1996rx}
D.~Cangemi, ``{Selfdual Yang-Mills theory and one loop like - helicity QCD
  multi - gluon amplitudes},'' {\em Nucl. Phys.} {\bf B484} (1997) 521--537,
\href{http://arXiv.org/abs/hep-th/9605208}{{\tt hep-th/9605208}}.

\bibitem{Cachazo:2004kj}
F.~Cachazo, P.~Svrcek, and E.~Witten, ``{MHV vertices and tree amplitudes in
  gauge theory},'' {\em JHEP} {\bf 09} (2004) 006,
\href{http://arXiv.org/abs/hep-th/0403047}{{\tt hep-th/0403047}}.

\bibitem{Gorsky:2005sf}
A.~Gorsky and A.~Rosly, ``{From Yang-Mills Lagrangian to MHV diagrams},'' {\em
  JHEP} {\bf 01} (2006) 101,
\href{http://arXiv.org/abs/hep-th/0510111}{{\tt hep-th/0510111}}.

\bibitem{Mansfield:2005yd}
P.~Mansfield, ``{The Lagrangian origin of MHV rules},'' {\em JHEP} {\bf 03}
  (2006) 037,
\href{http://arXiv.org/abs/hep-th/0511264}{{\tt hep-th/0511264}}.

\bibitem{Britto:2004ap}
R.~Britto, F.~Cachazo, and B.~Feng, ``{New recursion relations for tree
  amplitudes of gluons},'' {\em Nucl. Phys.} {\bf B715} (2005) 499--522,
\href{http://arXiv.org/abs/hep-th/0412308}{{\tt hep-th/0412308}}.

\bibitem{Britto:2005fq}
R.~Britto, F.~Cachazo, B.~Feng, and E.~Witten, ``{Direct proof of tree-level
  recursion relation in Yang-Mills theory},'' {\em Phys. Rev. Lett.} {\bf 94}
  (2005) 181602,
\href{http://arXiv.org/abs/hep-th/0501052}{{\tt hep-th/0501052}}.

\bibitem{Benincasa:2011kn}
P.~Benincasa and E.~Conde, ``{On the Tree-Level Structure of Scattering
  Amplitudes of Massless Particles},'' {\em JHEP} {\bf 11} (2011) 074,
\href{http://arXiv.org/abs/1106.0166}{{\tt 1106.0166}}.

\bibitem{Benincasa:2011pg}
P.~Benincasa and E.~Conde, ``{Exploring the S-Matrix of Massless Particles},''
  {\em Phys. Rev.} {\bf D86} (2012) 025007,
\href{http://arXiv.org/abs/1108.3078}{{\tt 1108.3078}}.

\bibitem{McGady:2013sga}
D.~A. McGady and L.~Rodina, ``{Higher-spin massless $S$-matrices in
  four-dimensions},'' {\em Phys. Rev.} {\bf D90} (2014), no.~8, 084048,
\href{http://arXiv.org/abs/1311.2938}{{\tt 1311.2938}}.

\bibitem{Bekaert:2015tva}
X.~Bekaert, J.~Erdmenger, D.~Ponomarev, and C.~Sleight, ``{Quartic AdS
  Interactions in Higher-Spin Gravity from Conformal Field Theory},'' {\em
  JHEP} {\bf 11} (2015) 149,
\href{http://arXiv.org/abs/1508.04292}{{\tt 1508.04292}}.

\bibitem{Barnich:1993vg}
G.~Barnich and M.~Henneaux, ``{Consistent couplings between fields with a gauge
  freedom and deformations of the master equation},'' {\em Phys. Lett.} {\bf
  B311} (1993) 123--129,
\href{http://arXiv.org/abs/hep-th/9304057}{{\tt hep-th/9304057}}.

\bibitem{Fotopoulos:2010ay}
A.~Fotopoulos and M.~Tsulaia, ``{On the Tensionless Limit of String theory, Off
  - Shell Higher Spin Interaction Vertices and BCFW Recursion Relations},''
  {\em JHEP} {\bf 11} (2010) 086,
\href{http://arXiv.org/abs/1009.0727}{{\tt 1009.0727}}.

\bibitem{Ponomarev:2016jqk}
D.~Ponomarev and A.~A. Tseytlin, ``{On quantum corrections in higher-spin
  theory in flat space},'' {\em JHEP} {\bf 05} (2016) 184,
\href{http://arXiv.org/abs/1603.06273}{{\tt 1603.06273}}.

\bibitem{Bern:2008qj}
Z.~Bern, J.~J.~M. Carrasco, and H.~Johansson, ``{New Relations for Gauge-Theory
  Amplitudes},'' {\em Phys. Rev.} {\bf D78} (2008) 085011,
\href{http://arXiv.org/abs/0805.3993}{{\tt 0805.3993}}.

\bibitem{Bern:2010ue}
Z.~Bern, J.~J.~M. Carrasco, and H.~Johansson, ``{Perturbative Quantum Gravity
  as a Double Copy of Gauge Theory},'' {\em Phys. Rev. Lett.} {\bf 105} (2010)
  061602,
\href{http://arXiv.org/abs/1004.0476}{{\tt 1004.0476}}.

\bibitem{Conde:2016vxs}
E.~Conde and A.~Marzolla, ``{Lorentz Constraints on Massive Three-Point
  Amplitudes},'' {\em JHEP} {\bf 09} (2016) 041,
\href{http://arXiv.org/abs/1601.08113}{{\tt 1601.08113}}.

\end{thebibliography}\endgroup
\bibliographystyle{utphys}
\end{document}